\documentstyle[12pt,epsfig]{article}
\def\beq{\begin{equation}}
\def\eeq{\end{equation}}
\def\beqa{\begin{eqnarray}}
\def\eeqa{\end{eqnarray}}

\def\ie{{\it i.e. }}

\def\jmp{{\it J. Math. Phys.}\ }
\def\pr{{\it Phys. Rev.}\ }
\def\prl{{\it Phys. Rev. Lett.}\ }
\def\pl{{\it Phys. Lett.}\ }
\def\np{{\it Nucl. Phys.}\ }
\def\mpl{{\it Mod. Phys. Lett.}\ }
\def\ijmp{{\it Int. Journ. Mod. Phys.}\ }

\def\cqg{{\it Class. Quantum Grav.}\ }
\def\aph{{\it Ann. Phys.}\ }

\def\grg{{\it Gen. Relativ. Grav.}\ }

\def\apj{{\it Ap. J.}\ }
\def\aj{{\it Astron. J.}\ }
\def\aa{{\it Astron. Astrophys.}\ }
\def\ncim{{\it Nuovo Cim.}\ }

\def\mnras{{\it Mon. Not. R. Ast. Soc.}\ }

\def\rmp{{\it Rev. Mod. Phys.}\ }

\def\ass{{\it Astr. and Space Sci.}\ }

\begin{document}
\def\bib#1{[{\ref{#1}}]}
\begin{titlepage}
\title{Quintessence without scalar fields}

\author{{Salvatore Capozziello\thanks{capozziello@sa.infn.it}, Sante Carloni\thanks{carloni@sa.infn.it},
Antonio Troisi\thanks{antro@sa.infn.it}}
\\ {\it Dipartimento di Fisica``E. R. Caianiello'', } \\
 {\it Universit\`{a} di Salerno, I-84081 Baronissi, Salerno,} \\
 {\it Istituto Nazionale  di Fisica Nucleare, sez. di Napoli},\\{\it Gruppo
Collegato di Salerno,}\\{ Via S. Allende-84081 Baronissi (SA),
Italy}}

\date{\today}

\maketitle

\begin{abstract}
The issues of quintessence and cosmic acceleration can be
discussed in the framework of theories which do not include
necessarily scalar fields. It is possible to define pressure and
energy density for new components considering effective theories
derived from fundamental physics like the extended theories of
gravity or simply generalizing the state equation of matter. Exact
accelerated expanding solutions can be achieved in several
schemes: either in models containing higher order curvature and
torsion terms or in models where the state equation of matter is
corrected by a second order Van der Waals terms. In this review,
we present such new approaches and compare them with observations.

\end{abstract}

\thispagestyle{empty} \vspace{20.mm}
 PACS number(s): 98.80.Cq, 98.80. Hw, 04.20.Jb, 04.50 \\

\vspace{5.mm}

\vfill

\end{titlepage}

\section{\large\bf Introduction}

One of the recent astonishing result in cosmology is the fact that
the universe is accelerating instead of decelerating along the
scheme of standard Friedmann  models as everyone has learned in
textbooks \cite{kolb,peebles,peacock}. Type Ia supernovae (SNe Ia)
allow to determine cosmological parameters probing the today
values of Hubble constant $H_0$ in relation to the luminosity
distance deduced from these stars used as standard candles
\cite{perlmutter,riess}.

For the red-shift $z\leq 1$, by the luminosity distance
$d_{L}\simeq H_0^{-1}[z+(1-q_0)z^2/2]$, the observations indicate
that the deceleration parameter is

\begin{equation}
\label{1}
 -1\leq q_0 < 0
\end{equation}
which is a clear indication for the acceleration.

 Besides, data coming from clusters of galaxies
at low red shift  (including the mass-to-light methods, baryon
fraction and abundance evolution) \cite{cluster}, and data coming
from the CMBR ({\it Cosmic Microwave Background Radiation})
investigation (e.g. BOOMERanG)\cite{boomerang} give observational
constraints from which we deduce the picture of a spatially flat,
low density universe dominated by some kind of non-clustered dark
energy. Such an energy, which is supposed to have dynamics, should
be the origin of cosmic acceleration.

If we refer to the density parameter, all these observations seem
to indicate:
\begin{equation}
\label{2} \Omega_{M}\simeq 0.3\,,\qquad \Omega_{\Lambda}\simeq
0.7\,, \qquad \Omega_{k}\simeq 0.0\,,
\end{equation}
where $\Omega_{M}$ is the amount of both non-relativistic baryonic
and non-baryonic (dark) matter, $\Lambda$ is the dark energy
(cosmological constant, quintessence,..), $k$ is the curvature
parameter of Friedmann-Robertson-Walker (FRW) metric of the form
\begin{equation}
\label{3}
 ds^2=dt^2-a(t)^2\left[\frac{dr^2}{1-k
r^2}+r^2d\Omega^2\right]\,,
\end{equation}
here $a(t)$ is the scale factor of the universe\footnote{The
deceleration parameter can be given in terms of densities
parameters. We have for FRW models
\begin{equation}
\label{nota}
q_0=-\frac{\ddot{a}a}{\dot{a}^2}=\frac{1}{2}(3\gamma+1)\Omega_{M}-\Omega_{\Lambda}\,,
\end{equation}
where $\gamma$ is the constant of perfect fluid state equation
$p=\gamma\rho$.}.

The relations (\ref{2}) come from the Friedmann-Einstein equation

\begin{equation}\label{4}
 H^{2}=\frac{8\pi G}{3}\rho-\frac{k}{a^{2}}
\end{equation}

\noindent where $H= \displaystyle{\frac{\dot{a}}{a}}$ is the
Hubble parameter. Specifically we have

\begin{equation}\label{5}
 \Omega_{M}={\frac{8\pi G_{N}\rho_{M}}{3H^{2}}},\quad
 \Omega_{\Lambda}={\frac{8\pi
 G_{N}\rho_{\Lambda}}{3H^{2}}}={\frac{\Lambda}{3H^{2}}},\quad
 \Omega_{k}=-{\frac{k}{(aH)^{2}}}
 \end{equation}

 \noindent for the various components of cosmic fluid. Dividing by $H^{2}$,
 we get the simple relation

 \begin{equation}
\label{6}
 1=\Omega_{M}+\Omega_{\Lambda}+\Omega_{k}
\end{equation}

\noindent and then, through observations Eq. (\ref{2}).\\
 After this discovery, a wide debate on
the interpretation of data has been developed, in particular about
the explanation of this new $\Lambda$ component, featured as a
negative pressure fluid, needed for the best fit of observations.
It is clear that one of the most important challenge for the
current and future researches in cosmology, beside the traditional
issues as large scale structure, initial conditions, shortcomings
of standard model and so on, is understanding the true fundamental
nature of such a dark
energy.\\
Several approaches has been pursued to this discussion until now.
Neglecting some interesting but more philosophical speculations
(i.e. so called Anthropic Principle see \cite{anthropic}), the
most interesting schemes can be summarized into three great
families: cosmological
constant, {\it time varying} cosmological constant and quintessence models. \\
The first approach is related to a time-independent, spatially
homogeneous component which is physically equivalent to the
zero-point energy of fields \cite{weinberg,ratra}. Such a scheme
is of fundamental importance since fixing the value of $\Lambda$
should
provide the vacuum energy of gravitational field. \\
Deriving such value is important also in the framework of the so
called No-Hair conjecture whose issue is to predict what will be
the fate of the whole universe as soon as, during the evolution,
$\Lambda$ will remain the only contributions to the cosmic energy
density\footnote{A more precise formulation of such a conjecture
is possible for a restricted class of cosmological models, as
discussed in \cite{wald}. We have to note that the conjecture
holds when every ordinary matter field, satisfies the dominant and
strong energy conditions \cite{hawking}. However it is possible to
find models which explicitly violate such conditions but satisfies
No--Hair theorem requests. This conjecture can be extended to the
case of time--varying cosmological constant \cite{capozziello}}
\cite{wald,hoyle}. This consideration implies shortcomings in such
an approach. The predicted value of $\Lambda$ parameter is very
tiny compared to typical values of zero-point energy coming from
particle physics
 (at least 120 orders of
difference), this discrepancy gives rise to the so called {\it
problem of cosmological constant}. Unfortunately, if we were able
to explain the small value of the cosmological constant, it would
not be enough. A second puzzle, likewise not easily explainable,
is the why matter density and cosmological constant component have
today comparable values ({\it coincidence problem}). In order to
solve these problems, many authors have investigated others
alternative schemes.\\ For example, a time-varying cosmological
constant has been introduced, in particular in relation to the
inflationary paradigm.\\
The main feature of inflationary approaches is the development of
a de Sitter (quasi-de Sitter or power law) expansion of early
universe. \\
To connect such an expansion to structure formation and to a
forthcoming de Sitter epoch (as requested by No--Hair conjecture)
\cite{guth,linde} we need a cosmological constant which value
acquired a great value in early epoch, underwent a phase
transition with a graceful exit and will to result in a small
remnant to ward the future \cite{capoz2}. This scheme provides a
mechanism to overcome the {\it{coincidence problem}}. \\
In fact by a dynamical component, it could be more natural to
achieve comparable values of energy density between cosmological
fluids at a given epoch. \\
  The straightforward generalization is to consider an inhomogeneous and un-clustered
  cosmological component. From this view point we arrive to
 the definition of {\it quintessence} \cite{steinhardt}, which is a time-varying
 spatially inhomogeneous fluid with negative pressure.\\
 It is clear that these three approaches are strictly
 linked each other. Constant vacuum energy density, quintessence
 and time varying cosmological constant converge to the limit $\gamma\rightarrow
 -1$, since inhomogeneities are not revealed for dark energy
 component at the observable scales. All the authors assume quintessence
  as just a time-varying cosmological component.
In summary all the above schemes claim for an {\it ingredient}
which is a non-clustered energy counterpart which we need to
explain observations (in particular acceleration of cosmic fluid).\\
In order to solve the {\it cosmological constant} and {\it
coincidence problem}, people widely take into account scalar
fields which are a form of matter capable of giving rise to a
state equation with negative pressure. The key ingredient of such
a dynamics is the form of interaction potential which several
times is phenomenological and unnatural since
it cannot be related to some fundamental effective theory.\\
 In this review, we want to show how quintessential scheme can be
achieved also without considering scalar fields as usually
discussed in literature.\\
First, we will show how an accelerated behaviour of cosmic fluid
can be achieved assuming a more physically motivated equation of
state for matter:  in particular taking into account second order
terms of Van der Waals form. This approach seems very intriguing
in relation to the possibility of mimicking some effects, as phase
transitions, which
characterize standard matter behaviour. \\
From another point of view, we investigate the extended theories
of gravity, that is theories of gravity which generalize the
standard Hilbert-Einstein scheme. In this setting, it is possible
to obtain a cosmological component by a geometrical approach,
relating its origin either to effective terms of quantum gravity
(higher order theories of gravity) or to, for example, effective
contributions of matter-spin interaction which modify the space-time geometry
(torsion).\\
 \\
The paper is organized as follows. Sect.2 is devoted to a review
of principal results coming from observational investigations as
supernovae, cosmic microwave background and Sunyaev-Zeldovich
method for clusters of galaxies. The most peculiar evidence coming
from the bulk of
data is the fact that today observed universe is accelerating.\\
In Sect.3, we outline the scheme of quintessence in the framework
of scalar field approach.\\
Sect.4 is devoted to the Van der Waals quintessence. It is
interesting to note that the only request of taking into account a
more physically motivated equation of state
gives rise to the acceleration, matching the observations.\\
In Sects.5 and 6, we take into account curvature and torsion
quintessence considering the presence of geometrical terms into
the Hilbert-Einstein action. Such ingredients are essential every
time one takes into account effective quantum corrections to the
gravitational field \cite{odintsov}.\\
Sec.7 is devoted to the discussion of results, conclusions and
further perspectives of our approach.

\section{\large\bf What the observations really say}

In this section we give a summary of the present status of
observations which claim for the presence of some form of
cosmological dark energy. However the picture which we present is
far from being exhaustive but it gives an indication of the
problem from several point of view.
\\ \\

\subsection{\normalsize\bf The data from SNeIa}

The most impressive result of current observational cosmology has
been obtained by the study of supernovae of type Ia (SNeIa). The
phase of supernova is the product of collapse of super-massive
stars which degenerate in a very powerful explosion capable of
increasing the magnitude of the star to values similar to those of
the host galaxy. The physics of this late stellar phase is still
matter of debated \cite{puy}, however it is possible to classify
supernovae in some types in relation to the spectral emission.
Type Ia indicates supernovae originated by white dwarf carbon and
oxygen--rich degenerated through the interaction in binary
systems. These supernovae are interesting from a cosmological
point of view since it has been found a characteristic
phenomenological relation between the amplitude of the light-curve
and the maximum of luminosity $(\it{Phillips-relation})$. In this
way, they can be considered as good standard candles. This feature
has an immediate cosmological relevance: such kind of supernovae
can be detected at
high red-shifts (e.g. till $z \simeq 1$).\\
In order to see how supernovae work as standard candles we have to
give some notions related to the luminosity distance \cite{peebles,goobar}.\\
It is possible to define the luminosity distance $d_{L}$ of an
astrophysical object as a function of cosmological parameters. In
tha case of a FRW, spatially flat, metric we have:

\begin{equation}\label{7}
d_L(z) = (1 + z) \int_{0}^{z}{dz' [ \Omega_{M} (1 + z')^3 +
\Omega_{\Lambda} ]^{-1/2}} \ .
\end{equation}

\noindent Now, using the well known magnitude-red-shift relation

\begin{equation}\label{8}
\mu(z)=5\log\frac{c}{H_{0}}d_{L}(z)+25,
\end{equation}

\noindent we obtain a measure of the distance modulus $\mu$ in
terms of
cosmological parameters.\\
Distance modulus can be obtained from the observations of SNeIa.
In fact, the apparent magnitude $m$ is measured, while the
absolute magnitude $M$ may be deduced from the intrinsic
properties of these stars and some adjustment of the Phillips
relation as the Multi-Color Light-curve Shape (MLCS) method
\cite{filippenko}. The distance modulus is simply $\mu=M-m$.
Finally the red-shift $z$ of the supernova can be determined
accurately from the host galaxy spectrum. At this point, observing
a certain sample of supernovae \cite{perlmutter, riess} it has
been possible to fit different cosmological models which give rise
to different luminosity
distances.\\
The two groups SCP ({\it Supernovae Cosmology Project})
\cite{perlmutter} and HZT ({\it High z Search Team}) \cite{riess}
found that distant supernovae are significantly dimmer (nearly
half a magnitude), with respect to a sample of nearby supernovae,
than it would expected in a cosmological model with
$\Omega_{M}=1$, (that is the standard cold dark matter
model).\\
Referring to the best fit values, the SCP group suggests a
universe with
 \begin{equation}\label{9}
0.8{\Omega_{M}}-0.6{\Omega_{\Lambda}}\simeq{-0.2}\pm{0.1},
\end{equation}
which gives for a flat model with
\begin{equation}\label{10}
{\Omega_{M}}{\simeq{0.28}}, \quad {\Omega_{\Lambda}}\simeq{0.72}.
\end{equation}
On the other side, HZT constraints are compatible with the ones of
SCP; in fact also in this case the best-fit value for the flat
case is ${\Omega_{M}}\simeq{0.28}$.\\ From these estimates of
matter density parameter, it results that a spatially flat
universe should be filled by a dark cosmological component
with a negative pressure and density parameter $\simeq{0.7}$.\\
As consequence it is easy to obtain a negative value for the
deceleration parameter discussed above.\\ Other results coming
from these observations are a best fit value for
$H_{0}\simeq{65}kmsec^{-1}Mpc^{-1}$ and an estimate for the age of
universe of about 15Gyr.

\subsection{\normalsize\bf Cosmic microwave background radiation}

Another class of astrophysical observations with great relevance
in cosmology is the study of CMBR. This radiation has been
produced at recombination, when matter is become transparent to
light and because cosmic expansion has cooled the universe at a
today temperature of about $T_{0}=2.73K$. The CMBR can be assumed
coming from a far spherical shell around us, the so called
{\it{last scattering surface}} and can be considered, in a first
approximation, homogeneous and isotropic.\\
Now, from theories of structure formation we know that clustered
matter is the result of evolution of primordial perturbations
acting as seeds. So, it seems obvious to think to an influence of
such perturbations on the background radiation. We expect primary
and secondary anisotropies. The first ones should be produced just
at recombination as an imprint of inhomogeneities on the last
scattering surface (Sachs-Wolfe effect, intrinsic adiabatic
fluctuations, etc..) \cite{kolb,peebles}. The secondary ones are,
instead due to the effect of scattering processes along the line
of sight between the surface of last
scattering and the observer.\\
By considering only primary fluctuations, we can deal with the
matter of early universe as a fluid of photons and baryons. We
have a competition of gravity and radiation pressure effects in
this fluid which implies the setting up of acoustic oscillations.
At the decoupling, these oscillations have been frozen out in the
CMBR and, today, we detect them as temperature anisotropies in the
observed sky. The relevance of such studies is related to the
possibility of linking the amplitude of fluctuations, and in
particular their power spectrum, to the spatial geometry of the
universe. It can be shown that angular scale or, equivalently, the
multipole momentum {\it l} of the first acoustic peak of spectrum
can be expressed in terms of spatial curvature. It is possible to
show that the relation \cite{maxima}
\begin{equation}\label{11}
{\it{l}}\approx\frac{200}{\sqrt{1-\Omega_{k}}}
\end{equation}
holds. It gives the first multipole as a function of energy
density of spatial curvature. \\
In 1992, the COBE satellite showed for the first time that such
fluctuations  in the CMBR are really existing. Starting from such
result, others surveys have been performed with the aim of probing
thermal fluctuations in the sky with a best sensibility. These
experiments, COBE-DMR ({\it COsmic Background
Explorer-Differential Microwave Radiometer}) \cite{cobe-dmr},
BOOMERanG ({\it Balloon Observations Of Millimetric Extragalactic
Radiation and Geophysics})\cite{boomerang}, MAXIMA ({\it
Millimeter Anisotropy eXperiment IMaging array })\cite{maxima},
have indicated for {\it l} the value $\approx{197\pm 6}$
(BOOMERanG) and $\approx{220}$ (MAXIMA). The estimate of spatial
curvature density parameter obtained combining the different
results is $ {\Omega_{k}\approx{0.11\pm{0.07}}}$
\cite{cobe-boom-maxim}. It must be stressed that such a number has
been obtained taking into account a value of Hubble parameter
which is deduced
thanks to the supenovae observations.\\
Previous results for $\Omega_{k}$ were slightly different from
zero, but it is possible to reduce them near to zero with a
reasonable agreement to the others cosmological parameters'
estimates \cite{cobe-boom-maxim}. In relation to this outcome, one
can say that the best-fit CMBR results give a picture of the
universe that is a spatially flat manifold. Such a result
 is in agreement with the  case of SNeIa surveys.

\subsection{\normalsize\bf Others observational approaches}

In addition to the investigations of CMBR and SNeIa surveys
several astrophysical observations can have cosmological
relevance, in order to probe structure
and dynamics of cosmic fluid.\\
Among these ones, weak gravitational lensing is a very active
research area which is giving extremely interesting result
\cite{bartelmann}.
\\ The relevance of lensing as a
cosmological probe is linked to the fact that different
cosmological models give different probabilities for the light
coming from very distant object to meet gravitational lenses on
their walk. This fact can be used to test models with
observations, in fact, as seen for the supernovae case, the
luminosity distance depends on the cosmological parameters as the
cosmic volume at a specific red-shift. As a result, the counting
of apparent density of observed objects, whose actual value is
assumed as known, provides a test to probe the universe components
\cite{carroll}. Considering a flat universe, the probability of a
source, at red-shift {\it z}, to be lensed, relative to the case
of an Einstein-de Sitter universe
(${\Omega_{M}=1,\Omega_{\Lambda}=0}$) is:
\begin{equation}\label{12}
P_{lens}=\frac{15}{4}[1-(1+{\it z^{*}})^{-1/2}]^{-3}\int_{1}^{\it
a_{*}}da\frac{H_{0}}{H(a)}\left[\frac{{d_{A}(0,a)}{d_{A}(a,a_{*})}}{d_{A}(0,a_{*})}\right],
\end{equation}
where $d_{A}=(1+z)^{-2}d_{L}$ is the angular distance and
$a_{*}=(1+z_{*})^{-1}$ corresponds to the redshift $z_{*}$. It
must be underlined that such observational approach is frustrated
by several uncertainties due to evolution, extinction and so on.
Taking into account such bias sources, it is possible to obtain an
upper limit for cosmological energy density component
$\Omega_{\Lambda}$ which is $\leq 0.7$
\cite{kochanek,falco,chiba}.\\ We can cite also other recent
studies conducted by weak lensing approach: in these cases the
amplification is dependent on cosmological models and the
data indicate always a nonzero $\Lambda-term$ \cite{durrer}. \\
Other important tests for cosmological parameters are related the
evaluation of $\Omega_{M}$. Studies on $\Omega_{M}$ have provided
values ranging between 0.1 and 0.4. A result drastically larger
than the density parameter for the baryon matter as inferred from
nucleosynthesis
$\Omega_{bar}\approx{0.04}$ \cite{schramm,burles}.\\
Typically the estimates of $\Omega_{M}$ are performed {\it
weighting} the mass of galaxy clusters in relation to their
luminosity and extrapolating the results to the whole universe.
Adopting this strategy, and using the virial theorem for the
dynamics of clusters, a value of $\Omega_{M}=0.2\pm{0.1}$ is
obtained \cite{bahcall}. Also others collaborations, which weight
clusters using gravitational lensing of background galaxies,
arrive to similar results
\cite{smail}.\\
Among the many approaches pursued to evaluate the whole content of
matter in the universe, there is the measurement of total mass
relative to the baryon density. In this case we need the value of
baryon density which implies the analysis of intracluster gas, the
X-ray emission and the Sunyaev-Zeldovich effect ({\it SZE}). These
measurements provide $\Omega_{M}=0.3\pm{0.1}$ \cite{carlstrom}
consistent to the estimates obtained by different approaches.
Tests performed to evaluate $\Omega_{M}$ are very important since
they establish an upper limit to this parameter and give
fundamental information in order to obtain indirect estimates of
other parameters as in the cases of CMBR and SNeIa surveys.
 \\The SZE effect
 and the thermal bremsstrahlung (X-ray brightness data) for
galaxy clusters represent intriguing astrophysical tests with
several cosmological implications. In fact, distances measurements
using SZE and X-ray emission from intracluster medium, are based
on the fact that such processes depend on different combinations
of some parameters of clusters (e.g. see \cite{birk} and
references therein).  We recall that the SZE is a result of the
inverse Compton scattering between CMBR photons and hot electrons
of intracluster gas. The photon number is preserved, but photons
gain energy and so   a decrease of temperature is generated in the
Rayleigh-Jeans region of black-body spectrum; on the other hand an
increment is obtained in the Wien region. As it is well known, the
Hubble constant $H_{0}$ and the density parameters can be
constrained by means of the angular distances; it is possible to
use the SZE and thermal bremsstrahlung to estimate such distances.
For a flat model with $\Omega_{M}=0.3, \ \ \Omega_{\Lambda}=0.7$,
we have $H_{0}=63\pm{3}km sec^{-1}Mpc^{-1}$ while for an open
universe with $\Omega_{M}=0.3$, it has been obtained
$H_{0}=60\pm{3}$
\cite{rephaeli}.\\
There are many other approaches able to provide indications on
cosmological parameters, however in this section we have just
given a brief review of current status of observational results,
avoiding of debating about errors, a relevant issue for which we
remand to the literature.\\
Besides, some of the above observational approaches and methods
will be furtherly discussed below in connection to the test of our
models with observations.

\section{\large\bf The Quintessence Approach. Are scalar fields
strictly necessary to get acceleration?}

The above observational results lead to the straightforward
conclusion that cosmic dynamics cannot be explained in the
traditional framework of standard model \cite{weinberg} so that
further ingredients have to be introduced into the game.\\
Several evidences suggest that, besides the four basic elements of
cosmic matter-energy content, namely: baryons, leptons, photons
and cold dark matter, we need a {\it fifth} element in order to
explain apparent acceleration on extremely large scales. As we
said above, under the standard of {\it quintessence}, we can enrol
every ingredient capable of giving rise to such an acceleration.
In other words, the old {\it four element cosmology} has to be
substituted by a new {\it five elements cosmology}.\\
In this section, we will outline the new {\it standard lore} which
was born in the last five or six years. The key element of such a
new scheme is the fact that a scalar field can give rise to both
the accelerated behaviour of cosmic fluid and the bulk of
unclustered dark energy.\\
 Let us start from the cosmological Einstein-Friedmann equations
 which can be deduced from gravitational field equations when a
 FRW metric of the form (\ref{3}) is imposed. We have:

 \begin{equation}\label{13}
 \frac{\ddot a}{a}=-\frac{1}{6}(\rho+3p),
\end{equation}

\begin{equation}\label{14}
\left(\frac{\dot a}{a}\right)^{2}+\frac{k}{a^{2}}=\frac{1}{3}\rho,
\end{equation}

\begin{equation}\label{15}
\dot{\rho}+3\left(\frac{\dot a}{a}\right)(\rho+p)=0\, ,
\end{equation}

\noindent where Eq.(\ref{13}) is the Friedmann equation for the
acceleration of scale factor $a(t)$, Eq.(\ref{14}) is the energy
constraint and Eq.(\ref{15}) is the continuity equation deduced
from Bianchi contracted identities. We are using physical (Planck)
units where $8 \pi G=\hbar=k_{B}=1$ unless otherwise stated. The
source of these equations is a perfect fluid of standard matter
where $\rho$ is the matter-energy density and $p$ is the pressure
of a generic fluid.\\
A further equation is
\begin{equation}\label{16}
p=\gamma \rho , \end{equation}

\noindent which is the state equation which relates pressure and
energy-density. In standard cosmology
\begin{equation}\label{17}
0 \leq \gamma \leq 1,
\end{equation}

\noindent which is the so called Zeldovich interval where all the
forms of ordinary matter fluid are enclosed ($\gamma =0$ gives
"dust", $\gamma=\displaystyle \frac{1}{3}$ is "radiation",
$\gamma=1$ is "stiff matter").\\
In standard units, it is $\gamma=\displaystyle
\frac{c_{s}^{2}}{c^{2}}$ where $c_{s}$ is the sound speed of the
given fluid which cannot exceed light speed. Introducing the
relation (\ref{16}), with the constraint (\ref{17}), into
Eq.(\ref{13}), gives:

\begin{equation}\label{18}
\rho+3p=\rho(1+3\gamma)>0\, ,
\end{equation}

\noindent from which $\ddot a<0$ and then cosmological system
cannot accelerate in any way. To get acceleration, the constraint
(\ref{17}) has to be relaxed so that a fluid of some exotic kind
has to be taken into account. In other words, if the universe is
found to be accelerating, there must be a matter-energy component
$\rho_{Q}$ with negative pressure $p_{Q}$ such that

\begin{equation}\label{19}
\rho_{tot}+ 3p_{tot} <0 \Longrightarrow p_{tot} <
 - \frac{\rho_{tot}}{3}<0\, ,
\end{equation}

\noindent where $\rho_{tot}$ is the total matter-energy of the
system. Obviously, $\gamma$ loses its physical interpretation of
the squared ratio between speeds of sound and light. Since
$\rho_{Q}+p_{Q} \geq0$ for any physically plausible negative
pressure component and thanks to the positive energy conditions,
$\rho_{Q}>0$, we must have, at least,

\begin{equation}\label{20}
\rho_{Q}>\frac{\rho_{tot}}{3}\Rightarrow \ddot{a}>0\, ,
\end{equation}

\noindent where we are assuming all other components with
non-negative pressure. In order to match all the above conditions,
we use, as a paradigm, the evolution of a scalar field $Q$ ({\it
quintessence scalar field}) slowly rolling down its potential
$V(Q)$. In FRW space-time, we can define

\begin{equation}\label{21}
\rho_{Q}=\frac{1}{2}\dot{Q}^{2}+V(Q)\, ,
\end{equation}

\noindent which is the energy density of $Q$,

\begin{equation}\label{22}
p_{Q}=\frac{1}{2}\dot{Q}^{2}-V(Q)\, ,
\end{equation}

\noindent which is the pressure. The evolution is given by the
Klein-Gordon equation

\begin{equation}\label{23}
\ddot{Q}+3H\dot{Q}+V'(Q)=0\, ,
\end{equation}

\noindent while the Hubble parameter, in the case of a spatially
flat space-time, is given by

\begin{equation}\label{24}
H^{2}=\left(\frac{\dot{a}}{a}\right)^{2}=\frac{1}{3}(\rho_{M}+\rho_{Q})\,
,
\end{equation}

\noindent where $\rho_{M}$ is the ordinary matter energy density.
A suitable state equation is defined by

\begin{equation}\label{25}
\gamma_{Q}=\frac{p_{Q}}{\rho_{Q}}=\frac{\frac{1}{2}\dot{Q}^{2}-V(Q)}{\frac{1}{2}\dot{Q}^{2}+V(Q)},
\end{equation}

\noindent and acceleration is obtained by the constraint

\begin{equation}\label{26}
-1\leq \gamma_{Q}<0.
\end{equation}

\noindent In this approach, quintessence is a time-varying
component with negative pressure constrained by Eq.(\ref{26}).
Vacuum energy density (i.e. cosmological constant) is quintessence
in the limit $\gamma_{Q}\rightarrow -1$. In order to give rise to
structure formation quintessence could be also spatially
inhomogeneous as some authors claim
\cite{steinhardt}.\\
A crucial role is played by the potential $V(Q)$ which has strict
analogies with inflationary potentials. Also in the quintessence
case, slow-roll is a condition in which the kinetic energy is less
than the potential energy. Such a mechanism naturally produces a
negative pressure. The key difference with inflation is that
energy-scale for quintessence is much smaller and the associated
time-scale is much longer compared to
inflation.\\
Several classes of potentials can give rise to interesting
behaviours. For example, "runaway scalar fields" are promising
candidates for quintessence. In this case, the potential $V(Q)$,
the slope $V'(Q)$, the curvature $V''(Q)$ and the ratios $V'/V$
and $V''/V$ have to converge to zero as $Q\rightarrow \infty$
(primes indicate derivative with respect to {\it Q}).\\
The so-called {\it tracking solution} \cite{steinhardt} are of
particular interest in solving the coincidence problem. They are a
sort of cosmological attractors which are obtained if the two main
conditions
\begin{equation}\label{27}
\gamma_{Q}<\gamma_{M}\ \ \ \ \ \
\Gamma\equiv{\frac{V''V}{V'}^{2}}>1
\end{equation}

\noindent holds. A specific kind of tracking solutions gives rise
to the {\it creeping quintessence} which occurs for tracker
potentials if initially $\rho_{Q}\gg \rho_{M}$. Examples of
potentials useful to get quintessence are

\begin{equation}\label{28}
V(Q)\sim Q^{-\alpha},\ \ \ \ V(Q)\sim \exp({1/Q})
\end{equation}

\noindent but several other families are possible \cite{rubano}.\\
To conclude, we can say that a mechanism capable of producing
accelerated expansion of cosmic flow can be obtained by
introducing phenomenological scalar fields. As in the case of
inflation, this is a paradigm which could be implemented also in
other ways (for example, let us remember the Starobinsky
inflationary scalaron \cite{starobinsky-scalaron}).\\
The main criticism which could be risen to the above approach is
that, up to now, no fundamental scalar field has been found acting
as a quintessential dark energy field. For example, there is no
analogous of Higgs boson for quintessence.\\
In the following, we propose other approaches with the aim to show
that scalar fields are not strictly necessary to get quintessence.
We will explore three possibilities:
\begin{itemize}
\item{The use of a better physically motivated equation of state
(Van der Waals equation)}. \item{The extension of Einstein-Hilbert
gravitational action including higher-order curvature invariant
terms}.\item{Considering torsion as a further source in
cosmological equations.}
\end{itemize}

\section{\large\bf First hypothesis: Van der Waals Quintessence}

\subsection{\normalsize\bf The model}

As previously said, the quintessential scheme can be achieved also
by taking into account more general equations of state without
adding any scalar field. The consideration we have done is the
following. A perfect fluid equation of state is an approximation
which is not always valid and which does not describe the phase
transitions between successive thermodynamic states of cosmic
fluids. In some epochs of cosmological evolution, for example at
equivalence, two phases could be existed together. In these cases,
a simple description by a perfect fluid equation of state could
not be efficient.

A straightforward generalization can be achieved by taking into
account the Van der Waals  equation of state which describes a two
phase fluid. Also in this case, we have an approximation but it
has interesting consequences on dynamics.

Let us construct such a {\it Van der Waals cosmology} \cite{CdMF}.

Instead of the equation of state $p=\gamma\rho$, we take into
account the Van der Waals equation \cite{rowlinson}
\begin{equation}
\label{29-vdw1}
 p=\frac{\gamma \rho}{1-\beta \rho}-\alpha\rho^2\,,
\end{equation}
which gives the above one in the limit $\alpha,\beta\rightarrow
0$. In standard units, the coefficients are
\begin{equation}\label{30-vdw2}
\alpha=3p_c v_c^2=3p_{c}\rho^{-2}_{c}\,,\qquad
\beta=\frac{1}{3}v_c={(3\rho_c)}^{-1}\,.
\end{equation}
where $p_c$, $\rho_{c}$ and $v_c$ are critical pressure, density
and volume respectively. It is worth stressing the fact that we
are using standard forms of matter as dust or radiation but the
relation between the thermodynamic quantities $p$ and $\rho$ is
more elaborate. The critical values are the indications that
cosmic fluids  change their phases at certain thermodynamic
conditions.

For the sake of simplicity, let us consider the case $k=0$
(FRW-flat cosmology) as strongly suggested by the CMBR data
\cite{boomerang,maxima,netterfield}. We have, from Eqs.(\ref{13}),
(\ref{14}), (\ref{15}):

\beqa
\dot{H}&=&-H^2-\frac{1}{6}\left(\rho+3p\right)\label{31-vdw3},\eeqa
\beqa \dot{\rho}&=&-3H\left(\rho+p\right)\label{32-vdw4}, \eeqa
\beqa H^2&=&\frac{1}{3}\rho\label{33-vdw5}.
 \eeqa

The relation between $\rho$ and $p$ is given by
 Eq.(\ref{29-vdw1}), whereas Eq.(\ref{33-vdw5}) is a constraint, so that
 the effective variables of the system are $\{H,\rho\}$ and the phase space is a plane.
 The singular points are given by the conditions
 \beq\label{34-vdw6}
 \dot{H}=\dot{\rho}=0\,,
 \eeq
 which, introduced into Eqs.(\ref{31-vdw3}) and (\ref{32-vdw4}), give

 \beq\label{35-vdw7}
 p=-\rho\,.
 \eeq

 \noindent Immediately a de Sitter solution $a(t)=a_0\exp H_0t$ is found.
 Otherwise $(H=0)$, we obtain the static solution $a=a_0$.

 The values of $\rho$ which give such a situation are

 \beq\label{36-vdw8}
 H_{1,2}=\sqrt{\frac{\rho_{1,2}}{3}}\,,\qquad \rho_{1,2}=\frac{(\beta+\alpha)\mp\sqrt{(\beta+\alpha)^2
 -4\alpha\beta(1+\gamma)}}{2(1+\gamma)}\,,
 \eeq
 so that we have two singular points at finite. They depends on
 the set of parameters $\{\alpha,\beta,\gamma\}$ which, as we saw
 above, are connected to the critical thermodynamical parameters
 $\{p_c,v_c,\rho_c\}$ of the fluid described by the Van der Waals
 equation. However, we discard the solutions
 \beq\label{37-vdw9}
 H_{1,2}=-\sqrt{\frac{\rho_{1,2}}{3}}\,,
 \eeq
 which have no physical meaning in the today observed universe.

 In order to investigate the nature of such singular points, we
 have to "linearize" the dynamical system and study
 the local Lyapunov stability \cite{hirsch}. The system reduces to

 \beqa\label{38-vdw10}
 \dot{H}&=& AH+B\rho\,,\\\label{39-vdw11}
 \dot{\rho}&=& CH+D\rho\,,
 \eeqa
 where
 \beqa\label{40-vdw12}
 A&=&-H_{1,2}\,,\\
 B&=&-\left[\frac{1}{6}+\frac{\gamma}{2(1-\beta\rho_{1,2})^2}-\alpha\rho_{1,2}
 \right]\,,\\
 C&=&-3\left(\rho_{1,2}+\frac{\gamma\rho_{1,2}}{1-\beta\rho_{1,2}}-
 2\alpha\rho_{1,2}\right)\,,\\
 D&=&-3H_{1,2}\left[1+\frac{\gamma}{(1-\beta\rho_{1,2})^2}-
 2\alpha\rho_{1,2}\right]\,,
 \eeqa
 where the constraints (\ref{29-vdw1}) and (\ref{33-vdw5}) have
 been taken into account. Since  $H_{1,2}$, $\rho_{1,2}$,
 $\alpha$, $\beta$, $\gamma$ are real number, we have no
 imaginary component. The singular points can be nodes or saddle
 points. Their stability depends on the sign of $A,B,C,D$. If
 $A,B,C,D<0$, the singular point is a stable node which attract
 the trajectories of phase space. If $A,B,C,D>0$, the point is an
 unstable node. If the coefficients $A,B,C,D$ are both positive
 and negative the singular point is a saddle. In the case of a
 stable node, the prescriptions of No Hair Conjecture are
 matched so that $\rho_{1,2}$ can be seen as effective
 cosmological constants. It is interesting to note that, in this
 case, such a cosmological constants are not imposed {\it by hands}
 but they are recovered from the critical values of the Van der
 Waals fluid, i.e. $p_c, v_c, \rho_c$.

 The singular points at infinite can be studied by taking into
 account the equation
 \beq
 \label{41-vdw13}
 \frac{d H}{d\rho}=\frac{H^2+\frac{1}{6}
 \left(\rho+\frac{3\gamma\rho}{1-\beta\rho}-3\alpha\rho^2\right)}{3H\left(
 \rho+\frac{\gamma\rho}{1-\beta\rho}-\alpha\rho^2\right)}\,,
 \eeq
 which is constructed by the dynamical system (\ref{31-vdw3}-\ref{33-vdw5})
 and the Van der Waals relation (\ref{29-vdw1}). By imposing
 the asymptotic behaviour $H\sim \rho^m$,
we get $H\sim \rho^{1/2}$ as it has to be. However, prescriptions
of No Hair Conjecture are recovered also in this case. It is
interesting to observe that we get two  values of cosmological
''constant" at finite and at infinite which, in general, are
different.

The condition to obtain quintessence, i.e. an accelerated
behaviour, is

 \beq\label{42-vdw14} \rho+3p<0 \eeq

which, in our specific case, corresponds to
\begin{equation}
\label{43-vdw15} \rho\left(1+ \frac{3\gamma }{1-\beta
\rho}-3\alpha\rho\right)<0\,.
\end{equation}In terms of the parameters of Van der Waals fluid, the
conditions
\begin{equation}
\label{44-vdw16} \rho>\frac{1}{\beta}\,,\qquad \frac{\beta
+3\alpha}{\alpha\beta}>0\,,\qquad \frac{(\beta +
3\alpha)^2}{12\alpha\beta}>(1-3\gamma)\,,
\end{equation}
have to hold at the same time in order to get a positive matter
density.

 The observational constraints on $\Omega_{\Lambda}$ are achieved
  in terms of parameters $\{\alpha,\beta,\gamma\}$ being

\beq\label{45-vdw17}
\Omega_{\Lambda}=\Omega_{\Lambda}(\alpha,\beta,\gamma)=\frac{(\beta+\alpha)\mp\sqrt{(\beta+\alpha)^2
 -4\alpha\beta(1+\gamma)}}{2\rho_{crit}(1+\gamma)}\simeq 0.7
\eeq

which gives

\begin{equation}\label{46-vdw18}
-1\leq q_0(\alpha,\beta,\gamma)<0\,.
\end{equation}
As usual,
\begin{equation}\label{47-vdw19}
\rho_{crit}=\frac{3H_{0}^2}{8\pi G}\simeq 2{\times} 10^{-29}g
cm^{-3}
\end{equation}
is the cosmological critical density.\\
Such a qualitative discussion can be made quantitative by
constraining Van der Waals cosmology with observations.

\subsection{\normalsize\bf Constraining Van der Waals quintessence by observation}

The three parameters $(\alpha, \beta, \gamma)$ or $(\rho_c, p_c,
\gamma)$ which feature Van der Waals scheme are not independent on
each other. Considering the situation at critical points, one
gets\,:

\begin{equation}\label{48-vdw20}
p_c = \frac{3}{8} \gamma \rho_c \ , \label{eq: pcrhoc}
\end{equation}
so that the number of independent parameters is now reduced to
two, which are $(\rho_c, \gamma)$. Using Eq.(\ref{48-vdw20}), we
may rewrite Eq.(\ref{29-vdw1}) as\,:

\begin{equation}\label{49-vdw21}
y = \frac{3 \gamma x}{3 - x} - \frac{9}{8} \gamma x^2
\end{equation}
having introduced the scaled variables $x = \rho/\rho_c$ and $y =
p/\rho_c$. \\
In this case, since in our model there is only one fluid, its
nowaday density has to be equal to the critical density, i.e.\,:

\begin{equation}\label{50-vdw22}
\rho(z = 0; \rho_c, \gamma) =  \rho_{crit} \ \rightarrow \ \rho_c
= \frac{3 H_0^2}{8 \pi G} \ \frac{1}{x_0(\gamma)} \ , \label{eq:
rhocgamma}
\end{equation}
(hereinafter quantities labelled with $0$ refer to the today
values, i.e. at redshift $z = 0$).\\
At this point an important fact has to be stressed. It is clear
that the r.h.s. of Eq.(\ref{49-vdw21}) can assume positive,
negative and null values so that an "effective" matter-energy
density could be defined. In order to remove such a "disturbing
concept", we will generalize the parameter $\gamma$ which in our
approach is not simply given by
$\gamma=\displaystyle{\left(\frac{c_{s}}{c}\right)^2}$ but it
fixes the relation between $p_{eff}$ and $\rho_{eff}$, two
effective quantities assigned by the Van der Waals equation
(\ref{29-vdw1}). With this consideration in mind $\gamma$ can
assume negative values.\\
Using Eqs.(\ref{48-vdw20}) and (\ref{50-vdw22}), we can completely
characterize the cosmic fluid by only one parameter in the
equation of state, in other words $\gamma$ become the only
independent parameter needed to describe the Van der Waals fluid.
To constrain the value of $\gamma$, we can use some mathematical
and physical considerations. Let us consider Eq.(\ref{15}). We can
rewrite it as a differential relation between the scaled density
$x$ and the scale factor $a$ taking into account the Van der Waals
relation:

\begin{equation}\label{51-vdw22}
\frac{dx}{dt} = -3 \ (x + y) \ \frac{1}{a} \ \frac{da}{dt} \
\rightarrow \ -3 \ \frac{da}{a} = \frac{dx}{x + y} \ .
\end{equation}
It can be immediately integrated to give\,:

\begin{equation}
-3 \ (1 + \gamma) \ln{\left ( \frac{a}{a_0} \right )} =
\ln{P_1(x)} + P_2(x) \arctan{P_3(x)} \label{52-vdw23}
\end{equation}
with\,:

\begin{equation}
P_1(x) =  \frac{x}{\sqrt{9 \gamma x^2 - (27 \gamma + 8) x + 24 (1
+ \gamma)}} \ , \label{53-vdw24}
\end{equation}

\begin{equation}
P_2(x) = \frac{11 \gamma - 8}{\sqrt{135 \gamma^2 + 432 \gamma -
64}} \ , \label{54-vdw25}
\end{equation}

\begin{equation}
P_3(x) = \frac{18 \gamma x - 27 \gamma - 8}{\sqrt{135 \gamma^2 +
432 \gamma - 64}} \ . \label{55-vdw26}
\end{equation}
Since the quantity on the left hand side of Eq.(\ref{52-vdw23}) is
real, so must be the right hand side. We can use this simple
mathematical condition to determine the range of significant
values for $\gamma$. To this aim, it is useful to divide the real
axis in three different regions bounded by the roots of the second
order polynomial which is the function under the square root in
$P_2(x)$ and $P_3(x)$. Excluding the value $\gamma = -1$ in order
to avoid the divergence of $a$ due to the term $(1 + \gamma)$ in
Eq.(\ref{52-vdw23}), we have\,:

\begin{displaymath}
\gamma \ \in \ ] \ -\infty, -3.34186 \ [ \ \cup \ [ \ -3.34186,
0.141859 \ ]
\end{displaymath}
\begin{equation}
\ \ \ \ \cup \ ] \ 0.141859, \ \infty \ [ \  . \label{56-vdw27}
\end{equation}
We will examine the three different regions in Eq.(\ref{56-vdw27})
separately to see whether we can further constraint the range for
$\gamma$ on the basis of physical considerations.

\begin{figure}
\begin{center}
\includegraphics[width=9.5cm]{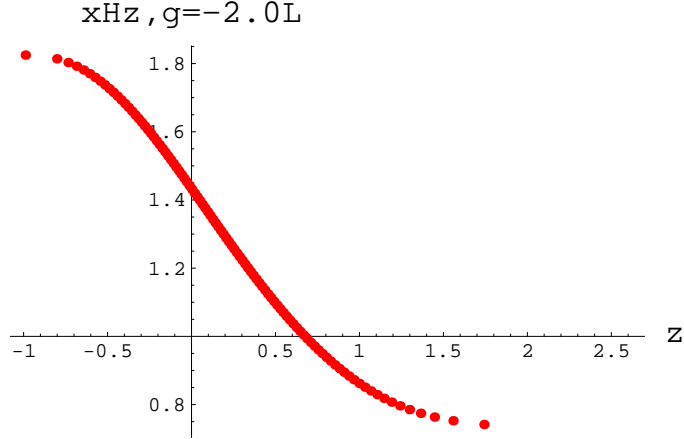}
\caption{\small The scaled density $x = \rho/\rho_c$ as function
of the redshift $z$ for the model with $\gamma = -2.0$. Note that
$x$ is confined between $(x_1, x_2)$ (with $x_1(\gamma = -2.0)
\simeq 1.823$ and $x_2(\gamma = -2.0) \simeq 0.731$) as discussed
in the text.}\end{center}
\end{figure}

\subsubsection{\normalsize\bf Models with $\gamma \in [ \ -3.34186, \ 0.141859 \
]$}

Let us start our analysis considering the second region in
Eq.(\ref{56-vdw27}). For values of $\gamma$ in this range, the
term under square root in $P_2(x)$ and $P_3(x)$ gets negative
values, but the right hand side is still a real quantity. To show
this, let us remember the two following mathematical relations
$\sqrt{b}=i\sqrt{b}$ ({\it b} real and negative), arctan${ib}=i$
arctanh${b}$ ({\it b} real), Eq.(\ref{52-vdw23}) may be rewritten
as:

\begin{equation}
-3 \ (1 + \gamma) \ln{\left ( \frac{a}{a_0} \right )} =
\ln{P_1(x)} + P'_2(x) {\rm arctanh{P'_3(x)}} \label{57-vdw28}
\end{equation}
with\,:

\begin{equation}
P'_2(x) = \frac{11 \gamma - 8}{\sqrt{- 135 \gamma^2 - 432 \gamma +
64}} \ , \label{58-vdw29}
\end{equation}

\begin{equation}
P'_3(x) = \frac{- 18 \gamma x + 27 \gamma + 8}{\sqrt{- 135
\gamma^2 - 432 \gamma + 64}} \ . \label{59-vdw30}
\end{equation}
Eq.(\ref{57-vdw28}) is correctly defined for $\gamma$, in the
range we are considering, provided that the logarithmic term is
real. To this aim, $x$ should be in the range $]\ x_1, \ x_2 \ [$,
being these latter the two roots of the square root term entering
$P_1(x)$, given as\,:

\begin{equation}
x_{1,2} = \frac{8 + 27 \gamma {\pm} \sqrt{64 - 432 \gamma - 135
\gamma^2}}{\gamma} \ . \label{60-vdw31}
\end{equation}
On the other hand, since $x$ is positive by its definition, we
must impose the constraint $x_{1,2} \ge 0$ which means that we
have to exclude all the values of $\gamma$ which lead to negative
values of $x_1$ or $x_2$. This constraint allows us to narrow the
second region in Eq.(\ref{56-vdw27}) which is now reduced to\,:

\begin{equation}
\gamma \in [ \ -3.34186, \ -1 \ [ \ . \label{61-vdw32}
\end{equation}
It is interesting to invert numerically Eq.(\ref{57-vdw28}) to get
$x(z; \gamma)$, being the red-shift $z$ related to the scale
factor as $a_0/a = 1 + z$. We plot the result for $\gamma = -2.0$
in Fig.\,1. This plot shows some interesting features. The scaled
energy density $x(z, \gamma)$ is bounded between two finite values
as a consequence of the condition imposed so that the logarithmic
term of Eq.(\ref{57-vdw28}) is correctly defined. It is
interesting to observe that the model predicts that the energy
density is an increasing function of time (and hence a decreasing
function of red-shift as in the super-inflationary models). The
two extreme values are $(x_1, x_2)$ that may be computed using
Eq.(\ref{60-vdw31}) once a value of $\gamma$ in the range defined
by Eq.(\ref{61-vdw32}) has been chosen. Furthermore, being $x$
constrained in the range $]x_1, x_2[$, we also get that the model
may describe the dynamics of the universe over a limited red-shift
range $(z_{min}, z_{max})$. This is not a serious shortcoming
since the Van der Waals equation of state we are using is only an
approximation (even if more realistic than the perfect fluid one)
of the true unknown equation of state. It is thus not surprising
that it may be applied only over a limited period of the universe
evolution. It is quite easy to show (both analytically and
numerically) that the lower limit of red-shift range is $z_{min} =
-1$ so that the model may be used also to describe the near future
evolution of universe. On the other hand, the upper limit is a
function of $\gamma$, in particular, $z_{max}$ is always larger
than $z \sim 12$ so that we can safely use the model to describe
dynamics of the universe over the red-shift range $(z \le 1)$
probed by SNeIa Hubble diagram. We will see, in next subsection,
that the situation will be different in the other red-shift ranges
defined by Eq.(\ref{56-vdw27}).

\begin{figure}
\centering \resizebox{9.5cm}{!}{\includegraphics{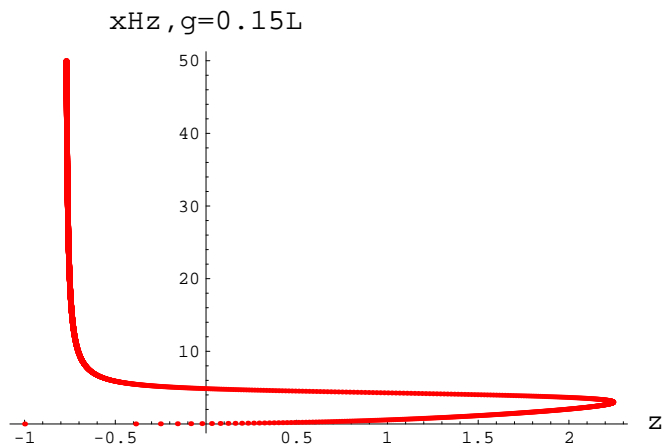}}
\caption{\small{The scaled density $x = \rho/\rho_c$ as function
of the redshift $z$ for the model with $\gamma = 0.15$. Note that
the apparent divergence on the left side is not a physical one,
but only an artifact of the scale used in the plot.}}
\end{figure}

\subsubsection{\normalsize\bf Models with $\gamma \not \in \ [ \ -3.34186, \
0.141859 \ ]$}

Let us turn now to the other two ranges defined in
Eq.(\ref{56-vdw27}). It is quite easy to show that the term under
square root, entering $P_1(x)$ in Eq.(\ref{52-vdw23}), has the
same sign of $\gamma$, if this parameter takes values in these
regions. In order to have a positive function, we have  to exclude
the region $] \ -\infty, \ -3.34186 \ [$ which we do not consider
anymore in the following so that we can now concentrate on the
range $] \ 0.141859, \ \infty \ [$.

We can further narrow it on the basis of physical considerations.
To this aim, let us invert numerically Eq.(\ref{52-vdw23}) to get
$x(z; \gamma)$. As an example, in Fig.\,2 the results for $\gamma
= 0.15$ are sketched. This plot shows two interesting features.
First, there is a limiting red-shift $z_{lim}$ such that, for $z
\ge z_{lim}$, the density is not defined. This fact simply states
that our model could describe the dynamical evolution of universe
only for red-shifts lower than $z_{lim}$. This fact could be
explained qualitatively observing that $x(z_{lim}; \gamma) = 3$
for every $\gamma$. For $\rho = 3 \rho_c$, we get $p\rightarrow
\infty$. It is quite obvious that the universe cannot evolve until
the pressure becomes finite (both positive or negative). Then, the
universe starts evolving, but the density may follow two different
tracks, decreasing or increasing with red-shift (that is
increasing or decreasing with time). We will come back to this
topic later.

We can observe that $z_{lim}$ is implicitly defined as function of
$\gamma$ by the relation $x(z_{lim}; \gamma) = 3$ and increases
with $\gamma$ faster and faster as $\gamma$ approaches the lowest
value in the range we are now considering.\\
In the following section, we will investigate whether our model is
able to reproduce the Hubble diagram of SNeIa. To this aim, the
minimal request is that the model describes the dynamical
evolution of matter density at least over the red-shift range
probed by SNeIa, i.e. until $z \sim 0.83$. However, future
observations (e.g. with the up to come SNAP satellite \cite{snap})
will be able to extend the upper limit of the range probed by
SNeIa until $z \sim 1.7$. A supernova at this red-shift has been
yet observed even if the estimated $z$ is very uncertain and its
apparent magnitude is likely to be seriously altered by
gravitational lensing effects.

Choosing a value of $\gamma$ very near to the lowest one will lead
to $z_{lim} > 1000$ so that our model could describe the universe
from the decoupling age up to now. However, the Van der Waals
equation of state is an approximation so that it should be not
surprising if the model works only on a limited range for $z$.
Thus, we impose an upper limit on the value of $\gamma$ by
requiring that $z_{lim}(\gamma_{max}) = 1.7$. This leads us to the
following range for $\gamma$\,:

\begin{equation}
\gamma \ \in \ ] \ 0.141859, \ 0.152370 \ [ \ . \label{62-vdw33}
\end{equation}
Before testing the models with $\gamma$ ranging in the interval
defined by Eqs.(\ref{61-vdw32}) and (\ref{62-vdw33}), we come back
again to Fig.\,2. This plot shows that, for $\gamma \in ] \
0.141859, \ 0.152370 \ [$, the universe starts evolving when $x
\ne 3$, but there are two possible evolutionary tracks. From a
mathematical point of view, this is a consequence of the intrinsic
non linearity of Eq.(\ref{52-vdw23}). However, the two curves are
clearly physically distinct since the density decreases or
increases with the red-shift according to which track is followed.
There is also another very important difference. Let us consider
Eq.(\ref{13}). If we ask for a today accelerating universe, we
have to impose the following constraint on the present day value
of the scaled density and pressure\,:

\begin{displaymath}
x_0 + 3 y_0 < 0 \ \rightarrow \ \frac{x_0}{8 \ (3 - x_0)}\left[ 27
\gamma x_0^2 - (81 \gamma + 8) x_0 + (72 \gamma + 24)\right] < 0 \
.
\end{displaymath}

For the values of $\gamma$ in the range defined by
Eq.(\ref{62-vdw33}), the second degree polynomial in this equation
is always positive definite, so that we get an accelerating
universe if\,:

\begin{displaymath}
3 - x_0 < 0 \ \longrightarrow \ x_0 > 3 \ .
\end{displaymath}
Fig.\,2 shows that the evolutionary track  to get a today positive
value of $\ddot{a}$ is the upper one, so that we arrive to the
conclusion that {\it the accelerated models are those with an
increasing density  with time}. Anyway, the result (from SNeIa)
that the universe is now in an accelerated phase is model
dependent so that, in the following, we will consider both models
and test them against the observations to discriminate among the
two evolutionary tracks.

\subsection{\normalsize\bf The Hubble diagram of SNeIa in the Van der Waals scheme}

It is well known that the use of astrophysical standard candles
provides a fundamental tool to test different cosmological models.
SNeIa are the best candidates to this aim since they can be
accurately calibrated and can be detected at enough high
red-shift. This fact allows to measure the SNeIa relation between
magnitude and red-shift (i.e. the Hubble diagram) at high enough
distances to discriminate among cosmological models. To this aim,
one can match a given model with the observed Hubble diagram,
conveniently expressed through Eq.(\ref{8}), which we write as\,:

\begin{equation}
\mu(z) = 5 \log{d_L(z)} + 25 \ , \label{63-vdw34}
\end{equation}
being $\mu$ as before the distance modulus and $D_L(z)$ the
luminosity distance calibrated with the Hubble constant. For our
cosmological models, it is\,:

\begin{equation}
d_L(z) = \frac{c}{H_0} \ \sqrt{x_0(\gamma)} \ (1 + z) \
\int_{0}^{z}{[ \ \rho(z') \ ]^{-1/2} \ dz'} \ . \label{64-vdw35}
\end{equation}
The distance modulus, as said, can be obtained from observations
of SNeIa. The apparent magnitude $m$ at the peak is measured,
while the absolute one $M$ may be deduced from template fitting or
using the Multi Color Light Curve Shape method
\cite{perlmutter,riess}. The distance modulus is then $\mu = m -
M$. Finally the redshift $z$ of the SNeIa is deduced from the host
galaxy spectrum or (with a larger uncertainty) from the supernova
spectrum directly. Our model can be fully characterized by two
parameters\,: the today Hubble constant $H_0$ and the value of
$\gamma$. We find their best fit values minimizing the $\chi^2$
defined as \cite{wang}\,:

\begin{equation}
\chi^2(H_0, \gamma) = \sum_{i = 1}^{N}{\frac{[ \
\mu_{i}^{theor}(z; H_0, \gamma) - \mu_{i}^{obs} \ ]^2}
{\sigma^{2}_{\mu_0,i} + \sigma^{2}_{mz,i}}} \label{65-vdw36}
\end{equation}
where the sum is over the data points. In Eq.(\ref{65-vdw36}),
$\sigma_{\mu_0}$ is the estimated error on the distance modulus,
while $\sigma_{mz}$ is the dispersion in the distance modulus due
to the uncertainty $\sigma_z$ on the measured red-shift. We
have\,:

\begin{equation}
\sigma_{mz} = \frac{5}{\ln{10}} \left (
\frac{1}{d_L(z)}\frac{\partial{d_L}}{\partial{z}} \right )
\sigma_z \ . \label{66-vdw37}
\end{equation}
Following Ref.\,\cite{wang}, we assume $\sigma_z = 200 \ {\rm km
s^{-1}}$ adding in quadrature $2500 \ {\rm km \ s^{-1}}$ for those
SNeIa whose red-shift is obtained from broad lines in their
spectrum. Note that $\sigma_{mz}$ depends on the parameters we
wish to determine so that we have to use an iterative procedure to
minimize the $\chi^2$.

The the SCP \cite{perlmutter} and HZT team \cite{riess} have
detected a quite large sample of high refshift $( \simeq 0.18 \div
0.83)$ SNeIa, while the Calan\,-\,Tololo survey \cite{hamuy} has
investigated the nearby sources. Using the data in
Refs.\,\cite{perlmutter, riess}, we have compiled a combined
sample of 85 SNeIa as described in Ref.\,\cite{wang}. We exclude 6
likely outlier SNeIa as discussed in Ref.\,\cite{perlmutter}, thus
ending with 79 SNeIa.

\begin{figure}
\centering
\resizebox{8.5cm}{!}{\includegraphics{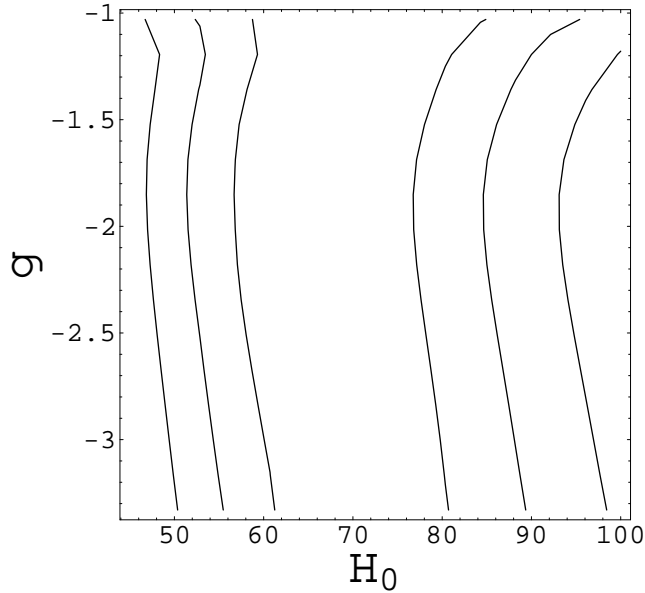}}
\caption{\small 1, 2 and 3\,-\,$\sigma$ confidence regions in the
$(H_0, \gamma)$ plane for the models with $\gamma \in \ ] \
-3.34186, \ -1 \ ]$.}
\end{figure}

Let us discuss first the results of the fit for the models with
$\gamma \in \ [ \ -3.34186, \ -1 \ [$.

As we can see in Fig.3 the models fit the data quite well. This is
an encouraging result which mean that the proposed equation of
state leads to a magnitude\,-\,red-shift relation which is in
agreement with the observed SNeIa Hubble diagram. This result is
further strengthened by the estimated value of the Hubble
constant. The best fit value is $H_0 = 71 \ {\rm km \ s^{-1} \
Mpc^{-1}}$, while the 68$\%$ confidence range turns out to be\,:

\begin{displaymath}
H_0 = (60\div 80) \ {\rm km \ s^{-1} \ Mpc^{-1}}\,,
\end{displaymath}
which is in very good agreement with the most recent estimates in
literature. For instance, the HST Key Project \cite{HSTkeyproj}
has calibrated different local distance estimators and has found
$H_0 = 72 \ {\pm} \ 8 \ {\rm km \ s^{-1} \ Mpc^{-1}}$. \\
It has to be stressed that the Hubble diagram of SNeIa (Fig.3)
gives no constraint on the $\gamma$ parameter. In fact, the best
fit value for this range turns out to be $\gamma = -1.03$, but the
$\chi^2$ changes less than 1$\%$ for $\gamma$ in the range defined
by Eq.(\ref{61-vdw32}). This is not an unexpected result since the
study of density gives a density $\rho(z,\gamma)$ almost
completely independent of $\gamma$ in the red-shift range probed
by the available data \cite{vdw}.

However, the dependence on $\gamma$ becomes significative for
higher red-shifts so that the prospects for using SNeIa to measure
the value of $\gamma$ are quite good as soon as SNeIa at $z \simeq
1.0 \div 1.7$ (a red-shift range which will be explored by the
dedicated SNAP satellite) will become available \cite{snap}.

\begin{figure}
\centering \resizebox{8.5cm}{!}{\includegraphics{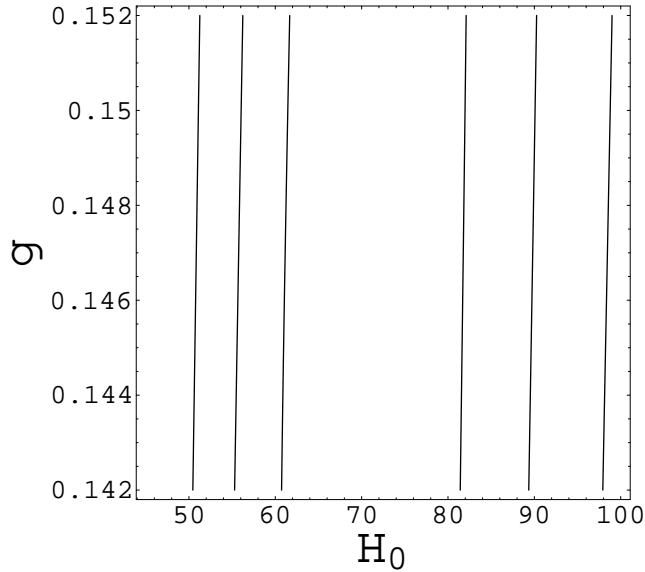}}
\caption{\small 1, 2 and 3\,-\,$\sigma$ confidence regions in the
$(H_0, \gamma)$ plane for the models with $\gamma \in \ ] \
0.141859, \ 0.151237 \ ]$ having the scaled energy density
increasing with the 5red-shift}
\end{figure}

The results of the fit for models with $\gamma$ in the range given
by Eq.(\ref{62-vdw33}) give also in this case that these models
are able to reproduce the observed SNeIa Hubble diagram. We obtain
$H_0 = 70 \ {\rm km \ s^{-1} \ Mpc^{-1}}$ as best fit value for
the Hubble constant, while the 68$\%$ confidence range is\,:

\begin{displaymath}
H_0 = (64\div 78) \ {\rm km \ s^{-1} \ Mpc^{-1}}\,,
\end{displaymath}
which is still in very good agreement with the most recent
estimates (see Fig.4). However, the Hubble diagram of SNeIa gives
no constraint on $\gamma$ parameter even for models with $\gamma$
in the range defined by Eq.(\ref{62-vdw33}). The best fit value
turns out to be $\gamma = 0.142$, but as in the previous case it
is not possible to give a reliable estimate of the uncertainty on
this parameter. This is again due to the almost complete
degeneracy of different models for $z \le 1$ .

On the other hand this degeneracy seems to be again eliminated
using SNeIa data in the red-shift range $z \simeq 1.0 \div 1.7$.

Finally, we have repeated the same analysis for  models with
density decreasing with time. Even if the fit is still possible,
the estimated $H_0$ is too low, the best fit value being $40 \
{\rm km \ s^{-1} \ Mpc^{-1}}$. This fact has led us to do not
consider anymore this class of models thus ending with the
surprising conclusion that the models which fit the data are those
with the energy density increasing with time.

\subsection{\normalsize\bf The age of the Universe in the Van der Waals scheme}

The  SNeIa--Hubble diagram fit has shown that several Van der
Waals models are able to reproduce the available data giving a
value of $H_0$ in very good agreement with the most recent
estimates. Furthermore, the matching with the data has allowed to
reject models with $\rho$ decreasing with time so that we do not
consider them anymore in the following analysis. However, we are
still not able to constrain the value of $\gamma$. To this aim, we
can use the available estimates of the age of the universe. As a
first step, we have to determine how the scale factor $a$ depends
on the time $t$. From Eq.(\ref{15}) and the choice
(\ref{49-vdw21}), we get\,:

\begin{equation}
t - t_{lim} = \frac{\sqrt{x_0(\gamma)}}{H_0} \int_{a_{lim}}^{a}{[
\ a' \sqrt{x(a')} \ ]^{-1} \ da'}\,, \label{67-vdw38}
\end{equation}
where $a_{lim}$ is the value of $a$ for the upper limit of
red-shift range where the model may describe the evolution of the
universe and $t_{lim}$ is the corresponding time. We have also
imposed $a_0 = 1$ for the sake of shortness. This relation cannot
be analytically inverted, but it is quite easy to invert
numerically.  Imposing $a = 1$ in Eq.(\ref{67-vdw38}), we  have an
estimate of $t_0 - t_{lim}$, but we do not know the value of
$t_{lim}$, i.e. the age of the universe at a red-shift when the
model starts to describe the dynamical evolution. On the other
hand, we can evaluate the quantity\,:

\begin{equation}
t_0 - t_{ref} = \frac{\sqrt{x_0(\gamma)}}{H_0}
\int_{a_{ref}}^{1}{[ \ a' \sqrt{x(a')} \ ]^{-1} \ da'}\,,
\label{68-vdw39}
\end{equation}
being $t_{ref}$ the age of the universe at a given red-shift
$z_{ref}$ with $a_{ref} = a(z_{ref})$. This approach allows to
constrain the values of $\gamma$, even if we do not know
$t_{lim}$, provided that we have an independent estimate of
$t_{ref}$. To this aim, we can use the age of an early-type high
red-shift galaxy which can be estimated in a way which does not
depend on the background cosmological model, but only on
astrophysical assumptions on the galaxy formation scenario and
stellar evolution. As a reference value, we use\,:

\begin{displaymath}
z_{ref} = 1.55 \ , \ t_{ref} = t_{1.55} > 3.5 \ {\rm Gyr}
\end{displaymath}
as first estimated in \cite{Dunlop} for the radio galaxy 53W091
\cite{Spinrad}. Different estimates of $t_0$ are present in
literature. Recent determinations of $t_0$ from the age of
globular clusters give $t_0 = 11.5 \ {\pm} \ 1.5$\,Gyr
\cite{CDKK98,K98}, while the analysis of the SNeIa Hubble diagram
suggests $t_0 = 14.9$ (for $H_{0}=63km\,s^{-1},Mpc^{-1}$)
\cite{perlmutter}. A multivariate analysis of combined BOOMERanG,
DASI and MAXIMA data on the CMBR anisotropy spectrum gives
estimates of $t_0$ ranging from $\sim 13$ to $\sim 16$\,Gyr
according to the former ones used in the maximum likelihood
estimation of the parameters \cite{Sievers}. We use $t_0 = 10 \div
16$\,Gyr so we get\,:

\begin{equation}
t_0 - t_{1.55} = 6.5 \div 12.5 \ {\rm Gyr} \label{69-vdw40}
\end{equation}
as a conservative limit on this quantity.

For each value of $\gamma$ in the range defined by
Eqs.(\ref{61-vdw32}) and (\ref{62-vdw33}) and for $H_0$ in the
corresponding 1\,$\sigma$ confidence interval, we can determine
the estimated range for $t_0 - t_{1.55}$. Imposing that this range
is contained within the range given by Eq.(\ref{69-vdw40}), we
find the constraint\,:

\begin{equation}
\gamma \ \in \ [ \ -1.95, \ -1.40 \ ] \ \label{70-vdw41}
\end{equation}
which we consider as our final estimate of the physically
acceptable range for $\gamma$. Note that the range in
Eq.(\ref{62-vdw33}) is completely cut out by this constraint since
the predicted values of $t_0 - t_{1.55}$ are much higher than the
(quite conservative) limits we have imposed.

\subsection{\normalsize\bf Discussion of results in Van der Waals cosmology}

We have investigated the possibility that a cosmological model
with  a Van der Waals  equation of state can give account of the
accelerating expansion of the universe as suggested by the SNeIa
data.

As a first step, starting from mathematical and physical
considerations, we have constrained  the effective parameter
$\gamma$ of the model to take values in the range $[\ -3.14186, \
-1 \ [$ $\ \cup \ ] \ 0.141859, \ 0.15237 \ ]$. In these regions,
the matter energy density can be both increasing or decreasing
with time as showed in Figs.\,1 and 2 and the quantity $x + 3 y$
results always negative which is consistent with an accelerated
phase of the universe evolution.

As a following step, we have fitted these models with the data of
SNeIa-Hubble diagram. The fitting procedure has lead us to
conclude that both the intervals of $\gamma$ are physically
significative, but the decreasing behaviour of energy density in
the range $]0.141859, 0.15237]$ has to be discarded since the
estimated Hubble constant is only $40 \ {\rm km \ s^{-1}\
Mpc^{-1}}$, in strong disagreement with results in literature. On
the other hand, the best fit values of the parameters turn out to
be\,:

\begin{displaymath}
\gamma = -1.09 \ , \ H_0 = 71 \ {\rm km \ s^{-1}\ Mpc^{-1}} \ ,
\end{displaymath}
for models with $\gamma \in \ ] \ -3.34186, \ -1 \ [$ and\,:

\begin{displaymath}
\gamma = 0.142 \ , \ H_0 = 70 \ {\rm km \ s^{-1}\ Mpc^{-1}},
\end{displaymath}
for models with $\gamma \in \ ] \ 0.14186, \ 0.15237 \ [$ and the
energy density increasing with time. The results for $H_0$ are in
perfect agreement with the estimates obtained with different
methods such as the calibration of local distance estimators
\cite{HSTkeyproj}, which is an evidence of the validity of our
models. Unfortunately, the very small variation of the energy
density with $\gamma$ in the red-shift range probed by available
data does not let us to discriminate in a useful way between
different values of this parameter. However, higher red-shift
SNeIa data (as the ones that will be collected by the next to come
SNAP satellite) will be able to break the degeneracy among the
different values of $\gamma$ since they will probe a red-shift
region where the energy density is more sensitive to the effective
parameter of the theory.

As a possible test to break the degeneracy among the values of
$\gamma$ with nowaday available data, we have estimated the
quantity $t_0 - t_{1.55}$, being $t_0$ the nowaday age of the
universe and $t_{1.55} = t (z = 1.55)$. For a fixed $\gamma$, we
estimate a range for $t_0 - t_{1.55}$ changing $H_0$ in the
corresponding 68\,$\%$ confidence region and compare it with the
conservative limit given in Eq.(\ref{69-vdw40}). This test allows
us to exclude the range in Eq.(\ref{62-vdw33}) since the predicted
values of $t_0 - t_{1.55}$ are much higher than 12.5\,Gyr so that
we turn out with a final estimate for $\gamma$ as\,:

\begin{displaymath}
\gamma \ \in \ [ \ -1.95, \ -1.40 \ ] \ .
\end{displaymath}
The analysis presented suggests that a cosmological model
characterized by a Van der Waals equation of state is a novel and
interesting approach to explain the accelerating expansion of the
universe. Indeed, for values of $\gamma$ in the range given by
Eq.(\ref{70-vdw41}), the models are able to fit quite well the
SNeIa Hubble diagram and predict an age of the universe consistent
with recent estimates obtained with completely different methods.
Another feature of the
 selected models is the  behaviour of the
energy density which is an increasing function of time (i.e.
decreasing with $z$ as shown in Fig.\,1) bounded between two
finite values. However, some models of super\,-\,inflation also
predict such a behaviour so that it should be interesting to see
whether a cosmological fluid with a Van der Waals equation of
state could be accomodated within the framework of these theories.

\section{\large\bf Second hypothesis: Curvature Quintessence}

\subsection{\normalsize\bf The model}

There is no {\it a priori} reason to restrict the gravitational
Lagrangian to a linear function of Ricci scalar $R$ minimally
coupled with matter \cite{francaviglia}. Furthermore, we have to
note that, recently, some authors have taken into serious
consideration the idea that there are no "exact" laws of physics,
but that Lagrangians of physical interactions are "stochastic"
functions with the property that local gauge invariances (i.e.
conservation laws) are well approximated in the low energy limit
with the properties that fundamental physical constants can vary
\cite{ottewill}. This scheme was adopted in order to treat the
quantization on curved spacetimes.  The result was that either
interactions among quantum fields and background geometry or the
gravitational self--interactions yield corrective terms in the
Einstein--Hilbert Lagrangian \cite{birrell}. Futhermore, it has
been realized that such corrective terms are inescapable if we
want to obtain the effective action of quantum gravity on scales
closed to the Planck length \cite{vilkovisky}. They are
higher--order terms in curvature invariants as $R^{2}$,
$R^{\mu\nu} R_{\mu\nu}$,
$R^{\mu\nu\alpha\beta}R_{\mu\nu\alpha\beta}$, $R\Box R$, or
$R\Box^{k}R$, or nonminimally coupled terms between scalar fields
and geometry as $\phi^{2}R$. Terms of this kind arise also in the
effective Lagrangian of strings and Kaluza--Klein theories when
the mechanism of dimensional reduction is working
\cite{veneziano}.

Besides fundamental physics motivations, all these theories have
acquired a huge interest in cosmology due to the fact that they
"naturally" exhibit inflationary behaviours and that the related
cosmological models seem very realistic \cite{starobinsky,la}.
Furthermore, it is possible to show that, via conformal
transformations, the higher--order and nonminimally coupled terms
({\it Jordan frame}) always corresponds to the Einstein gravity
plus one or more than one minimally coupled scalar fields ({\it
Einstein frame}) \cite{teyssandier,maeda,wands,conf,gottloeber} so
that these geometric contributions can always have a "matter-like"
interpretation. \\
Here we want to investigate the possibility that quintessence
could be achieved by extra-curvature contributions. We focus our
attention on fourth-order theories of gravity \cite{capcurv}.
\\
A generic fourth--order theory in four dimensions can be described
by the action
 \begin{equation}\label{71-curv1}
 {\cal A}=\int d^4x \sqrt{-g} \left[f(R)+{L}_{M} \right]\,{,}
 \end{equation}
 where $f(R)$ is a function of Ricci scalar $R$ and ${L}_{M}$
 is the standard matter Lagrangian density.
We are using physical units $8\pi G_N=c=\hbar=1$. The field
equations are

 \begin{equation}\label{72-curv2}
 f'(R)R_{\alpha\beta}-\frac{1}{2}f(R)g_{\alpha\beta}=
 f'(R)^{;\alpha\beta}(g_{\alpha\mu}g_{\beta\nu}-g_{\alpha\beta}g_{\mu\nu})+ \tilde{T}^{M}_{\alpha\beta}\,,
 \end{equation}
 which can be recast in the more expressive form
 \begin{equation}\label{73-curv3}
 G_{\alpha\beta}=R_{\alpha\beta}-\frac{1}{2}g_{\alpha\beta}R=T^{curv}_{\alpha\beta}+T^{M}_{\alpha\beta}\,,
 \end{equation}
 where
 \begin{equation}
 \label{74-curv4}
T^{curv}_{\alpha\beta}=\frac{1}{f'(R)}\left\{\frac{1}{2}g_{\alpha\beta}\left[f(R)-Rf'(R)\right]+
f'(R)^{;\alpha\beta}(g_{\alpha\mu}g_{\beta\nu}-g_{\alpha\beta}g_{\mu\nu})
\right\}
 \end{equation}
 and
 \begin{equation}
 \label{75-curv5}
 T^{M}_{\alpha\beta}=\frac{1}{f'(R)}\tilde{T}^{M}_{\alpha\beta}\,,
 \end{equation}
 is the stress-energy tensor of matter where we have taken into account
 the nonmnimal coupling to geometry. The prime indicates the derivative with respect to $R$.
 If $f(R)=R+2\Lambda$, the standard second--order gravity is
recovered.\\
It is possible to reduce the action to a point-like, FRW one. We
have to write
 \begin{equation}\label{76-curv6}
 {\cal A}_{curv}=\int dt {\cal L}(a, \dot{a}; R, \dot{R})\,{,}
 \end{equation}
where dot indicates derivative with respect to the cosmic time.
The scale factor $a$ and the Ricci scalar $R$ are the canonical
variables. This position could seem arbitrary since $R$ depends on
$a, \dot{a}, \ddot{a}$, but it is generally used in canonical
quantization \cite{schmidt,vilenkin,lambda}. The definition of $R$
in terms of $a, \dot{a}, \ddot{a}$ introduces a constraint which
eliminates second and higher order derivatives in action
(\ref{76-curv6}), and gives a system of second order differential
equations in $\{a, R\}$. Action (\ref{76-curv6}) can be written as
 \begin{equation}\label{77-curv7}
 {\cal A}_{curv}=2\pi^2\int dt \left\{ a^3f(R)-\lambda\left [ R+6\left (
 \frac{\ddot{a}}{a}+\frac{\dot{a}^2}{a^2}+\frac{k}{a^2}\right)\right]\right\}\,{,}
 \end{equation}
where the Lagrange multiplier $\lambda$ is derived by varying with
respect to $R$. It is
 \begin{equation}\label{78-curv8}
 \lambda=a^3f'(R)\,{.}
 \end{equation}
 The point-like Lagrangian is then
$$
 {\cal L}={\cal L}_{curv}+{\cal L}_{M}=a^3\left[f(R)-R
 f'(R)\right]+6a\dot{a}^2f'(R)+
$$
\begin{equation}
\label{79-curv9} \qquad +6a^2\dot{a}\dot{R}f''(R)-6ka
 f'(R)+a^3p_{M}\,,
 \end{equation}
 where we have taken into account also the fluid matter
 contribution which is, essentially, a pressure term
 \cite{quartic}.

 The Euler-Lagrange equations are
\begin{equation}
\label{80-curv10}
2\left(\frac{\ddot{a}}{a}\right)+\left(\frac{\dot{a}}{a}\right)^2+
\frac{k}{a^2}=-p_{tot}\,,
 \end{equation}
and
\begin{equation}
\label{81-curv11}
f''(R)\left[R+6\left(\frac{\ddot{a}}{a}+\frac{\dot{a}^2}{a}^2+\frac{k}{a^2}\right)\right]=0\,.
 \end{equation}
The dynamical system is completed by the energy condition
\begin{equation}
\label{82-curv12}
 \left(\frac{\dot{a}}{a}\right)^2+\frac{k}{a^2}=\frac{1}{3}\rho_{tot}\,.
\end{equation}

Combining Eq.(\ref{80-curv10}) and Eq.(\ref{82-curv12}), we obtain
 \begin{equation}
 \label{83-curv13}
 \left(\frac{\ddot{a}}{a}\right)=-\frac{1}{6}\left[\rho_{tot}+3p_{tot}
 \right]\,,
 \end{equation}
 where it is clear that the accelerated or decelerated behaviour
 depends on the r.h.s. How it is evident equations
 (\ref{80-curv10})-(\ref{83-curv13}) generalize in a very simply
 way the system (\ref{13})-(\ref{15}) for standard cosmological
 model.\\
 Now it holds:
 \begin{equation}
 \label{84-curv14}
 p_{tot}=p_{curv}+p_{M}\;\;\;\;\;\rho_{tot}=\rho_{curv}+\rho_{M}\,,
 \end{equation}
where we have distinguished the curvature and matter
contributions.

In fact by curvature-stress-energy tensor, we can define a
curvature pressure
\begin{equation}
\label{85-curv15}
p_{curv}=\frac{1}{f'(R)}\left\{2\left(\frac{\dot{a}}{a}\right)\dot{R}f''(R)+\ddot{R}f''(R)+\dot{R}^2f'''(R)
-\frac{1}{2}\left[f(R)-Rf'(R)\right] \right\}\,,
 \end{equation}
and a curvature density
\begin{equation}
\label{86-curv16}
\rho_{curv}=\frac{1}{f'(R)}\left\{\frac{1}{2}\left[f(R)-Rf'(R)\right]
-3\left(\frac{\dot{a}}{a}\right)\dot{R}f''(R) \right\}\,.
 \end{equation}

From Eq.(\ref{83-curv13}), the accelerated behaviour is achieved
if
\begin{equation}
\label{87-curv17} \rho_{tot}+ 3p_{tot}< 0\,,
\end{equation}
which means
\begin{equation}
\label{88-curv18} \rho_{curv}> \frac{1}{3}\rho_{tot}\,,
\end{equation}
assuming that all matter components have non-negative pressure.

In other words, conditions to obtain acceleration  depends on the
 relation
\begin{equation}
\label{89-curv19}
 \rho_{curv}+3p_{curv}=\frac{3}{f'(R)}\left\{\dot{R}^2f'''(R)+\left(\frac{\dot{a}}{a}\right)\dot{R}f''(R)
 +\ddot{R}f''(R)-\frac{1}{3}\left[f(R)-Rf'(R)\right]\right\}\,,
 \end{equation}
 which has to be compared with matter contribution. However, it
 has to be
 \begin{equation}
 \label{90-curv20}
 \frac{p_{(curv)}}{\rho_{curv}}=\gamma_{curv}\,,
 \qquad -1\leq\gamma_{curv}<0\,.
 \end{equation}
 The form of $f(R)$ is the main ingredient to obtain this {\it curvature
 quintessence}.
As simple choice in order to fit the above prescriptions, we ask
for solutions of the form
 \begin{equation}
 \label{91-curv21}
 f(R)=f_0 R^n\,,\qquad
 a(t)=a_0\left(\frac{t}{t_0}\right)^{\alpha}\,.
 \end{equation}
 The interesting cases are for $n\neq 1$ (being $n=1$ Einstein
 gravity) and $\alpha\geq 1$ (accelerated behaviour). Inserting
 Eqs.(\ref{91-curv21}) into the above dynamical system, we obtain the exact solutions
\begin{equation}
\label{92-curv22} \alpha=2\,;\qquad n=-1,3/2\,;\qquad k=0\,.
\end{equation}
In both cases, the deceleration parameter is
 \begin{equation}
 \label{93-curv23}
 q_0=-\frac{1}{2}\,,
 \end{equation}
 in perfect agreement with the observational results.

The case $n=3/2$ deserves further discussion. It is interesting in
conformal transformations from Jordan frame to Einstein frame
\cite{cqgconf,magnano} since it is possible to give explicit form
of scalar field potential. In fact, if
 \begin{equation}\label{94-curv24}
 \tilde{g}_{\alpha\beta}\equiv
 f'(R)g_{\alpha\beta}\,{,}\qquad
 \varphi=\sqrt{\frac{3}{2}}\ln f'(R)\,{,}
 \end{equation}
 we have the conformal equivalence of the Lagrangians
 \begin{equation}\label{95-curv25}
 {\cal L}=\sqrt{-g}\,f_0R^{3/2}\longleftrightarrow
 \tilde{{\cal L}}=\sqrt{-\tilde{g}}\left[-\frac{\tilde{R}}{2}+
 \frac{1}{2}\nabla_{\mu}\varphi\nabla^{\mu}\varphi-V_0\exp\left(
 \sqrt{\frac{2}{3}}\varphi\right)\right]\,,
 \end{equation}
 in our physical units. This is the so--called Liouville field
theory and it is one of the few cases where a fourth--order
Lagrangian can be expressed, in the Einstein frame, in terms of
elementary functions under a conformal transformation. Furthermore
this is a case in which quintessence can be achieved also by an
exponential potential.\\
It is possible to obtain the general solution \cite{lambiase}
 \begin{equation}\label{96-curv26}
 a(t)=a_0[c_4t^4+c_3t^3+c_2t^2+c_1t+c_0]^{1/2}\,{.}
 \end{equation}
\\
 The constants $c_i$ are combinations of the initial conditions.
Their particular values determine the type of cosmological
evolution. For example, $c_4\neq 0$ gives a power law inflation
while, if the regime is dominated by the linear term in $c_1$, we
get a radiation--dominated stage.\\
 These cases belong to an entire
 family of exact solutions for the quintessence curvature model.
 In fact solving Eqs. (\ref{83-curv13}) and (\ref{84-curv14}) in the
 limit in which matter can be neglected (this hypothesis can be considered well posed in
a toy model) and assuming for $f(R)$ and $a(t)$ behaviours like
(\ref{91-curv21}), we obtain the algebraic system for the
parameters $n$ and $\alpha$

\begin{equation} \label{97-curv27}
 \left\{ \begin{array}{ll} \alpha [\alpha(n-2)+2n^{2}-3n+1]=0 \\
 \\
\alpha[n^{2}+\alpha(n-2-n-1)]=n(n-1)(2n-1)\\
\end{array}
\right.
\end{equation}
\\

from which we have the solutions:
\begin{equation}
\label{98-curv28}
\begin{array}{cc} \alpha=0 \longrightarrow n=0,\,1/2,\,1\\ \\
\alpha=\displaystyle\frac{-2n^2+3n-1}{n-2}\,, \qquad
n\,\,.\end{array}
\end{equation}
\vspace{7mm}

 The cases for $\alpha=0$ give static models not interesting to
 discuss. The cases for $\alpha$ and $n$ generic deserve more
 attention since a class of cosmological quintessential behaviours
 can be found. In Fig.5 there is a sketch of the situation.

\begin{figure}
\begin{center}
 \includegraphics[width=10cm]{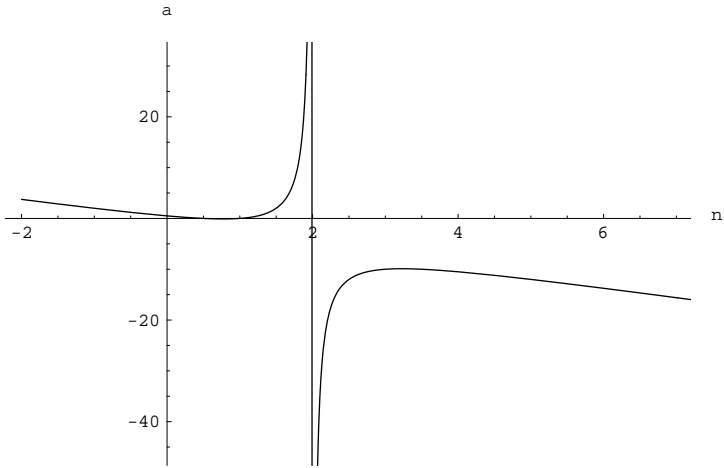}
  \caption{\small The plot shows the behaviour
of $\alpha$ and %$\gamma_{curv}$
 against $n$. It is evident a
region in which the power law of the scale factor is more than one
and correspondently the parameter $\gamma_{(curv)}$ of state
equation is negative .}\label{fig1}
\end{center}
\end{figure}

 The state equation for the class of solutions $n$ and
$\alpha=\displaystyle{\frac{-2n^2+3n-1}{n-2}}$ gives

\begin{equation}\label{99-curv29}
\gamma_{curv}=-\frac{6n^2-7n-1}{6n^2-9n+3}\,\,.
\end{equation}
\\

Accelerated behaviours for increasing scale factors are possible
only for $\gamma_{curv}< 0$ how requested for a cosmological fluid
with negative pressure. The whole approach seems intriguing in
relation
 to the possibility to get an accelerated phase of universe
 expansion as an effect of a higher-order curvature Lagrangian.
\\
 \\
 At this point, to obtain constraints on $n$
and to match with the observations we take into account data, in
particular we compare this model in relation to SNeIa results as
yet done for the Van der Waals case in (\cite{vdw} and
\cite{torsion}). To complete our analysis we have to evaluate also
the age prediction of curvature quintessence model as a
function of the parameter $n$.\\
In this sense, some results are present in literature where the
form of $f(R)$ is selected by the CMBR constraint \cite{hwang}.

\subsection{\normalsize\bf Matching Curvature Quintessence with observations }

\subsubsection{\normalsize\bf SNeIa matching}

In order to verify if such an approach has real perspective to be
physically acceptable, we match the theory with observational
data. In this way, we can constraint the parameters. \\
To this aim we consider the supernovae observations, in particular
we use the data in \cite{perlmutter, riess} as in the Van der Waals case.\\
Starting from these data, it is possible to perform a comparison
between the theoretically predicted expression of distance modulus
and its observational value. In this case the luminosity distance
can be expressed by the general expression:

\begin{equation}\label{100-curv30}
d_{L}=\frac{c}{H_{0}}(1+z)\int_{0}^{z}{\frac{1}{E_({\zeta})}}d{\zeta}
\end{equation}
\vspace{7mm}

\noindent where $E(\zeta)=\displaystyle\frac{H_{0}}{H}$ and $c$ is
the light speed having reintroduced the standard units.\\Starting
from Eq. (\ref{91-curv21}), the Hubble parameter, in terms of
redshift, is given by the relation

\begin{equation}\label{101-curv31}
H(z)=H_{0}(1+z)^{1/{\alpha}}
\end{equation}

\vspace{7mm}

\noindent where $\alpha$ is given in (\ref{98-curv28}) as a function of $n$.\\
The luminosity distance is now:

\begin{equation}\label{102-curv32}
d_{L}(z,H_{0},n)=\frac{c}{H_{0}}(1+z)\int_{0}^{z}(1+\zeta)^{-1/\alpha}d{\zeta}
\end{equation}
\\
and after the integration, we obtain

\begin{equation}\label{103-curv33}
d_{L}(z,H_{0},n)=\frac{c}{H_{0}}\left(\frac{\alpha}{\alpha-1}\right)(1+z)\left[
(1+z)^{\frac{\alpha}{\alpha-1}}-1\right]\,.
\end{equation}
\\
This expression is not defined for $\alpha=0,1$, so we have to
check the luminosity distance taking into account the existence of
such singularities. As a consequence the fit can be performed in
five intervals of $n$, that is:
$n<\displaystyle\frac{1}{2}(1-\sqrt{3})\,, \,\
\frac{1}{2}(1-\sqrt{3})<n<\frac{1}{2}\,,\,\ \frac{1}{2}<n<1\,,
1<n<\frac{1}{2}(1+\sqrt{3})\,,\,\ n>\frac{1}{2}(1+\sqrt{3})$.\\
Hubble parameter, as a function of $n$, shows the same trend of
$\alpha$ (Fig.5). We find that for $n$ negative, lower than -100,
it is strictly increasing while for $n$ positive, greater than
100, it is strictly decreasing. In relation to this feature, we
have tested $n$ for values ranging in these limits because, as we
shall see below, outside of this range the value of the age of
universe becomes manifestly not physically significant. The
results can be showed in the Table 1. From Table 1, in relation to
the best fit values of $H_0$ and $\chi^2$, we can only exclude the
range $\displaystyle\frac{1}{2}<n<1$ as physically not
interesting. In the other cases, the results give interesting best
fit values both for the Hubble parameter and the $\chi^2$,
indicating that such a model could represent a significant
theoretical background to explain SNeIa data.

\vspace{1cm}
\begin{table}
\begin{center}
 \begin{tabular}{|c|c|c|c|}
  \hline
  Range & $H_{0}^{best}$($km\,sec^{-1}Mpc^{-1}$)& $n^{best}$& $\chi^{2}$ \\
  \hline
  $-100<n<1/2(1-\sqrt{3})$ & $65$ & $-0.73$ & $1.003$  \\ \hline
  $1/2(1-\sqrt{3})<n<1/2$ & $63$ & $-0.36$ & $1.160$ \\ \hline
  $1/2<n<1$ & $100$ & $0.78$ & $348.97$ \\ \hline
  $1<n<1/2(1+\sqrt{3})$ & $62$ & $1.36$ & $1.182$ \\ \hline
  $1/2(1+\sqrt{3})<n<3$ & $65$ & $1.45$ & $1.003$ \\ \hline
  $3<n<100$ & $70$ & $100$ & $1.418$ \\ \hline
\end{tabular}
\end{center}
\caption{\small Results obtained fitting the curvature
quintessence model with SnIa data. First column indicate the range
of $n$ studied, column two gives the relative best fit $H_{0}$,
column three $n^{best}$, column four the $\chi^{2}$ index.}
\end{table}
\vspace{1cm}

We have to note that the best fit is completely degenerate with
respect to the $n$ parameter. This peculiarity indicates that
$\chi^{2}$ varies
slightly with respect to $n$ hindering the possibilities of constrain $n$.\\

\subsubsection{\normalsize\bf The age of universe in Curvature
Quintessence approach}

The age of the universe can be simply obtained from a theoretical
point of view if one has the value of the Hubble parameter. Now,
in curvature quintessence model from the definition of $H$, using
the relation (\ref{91-curv21}), we have:

\begin{equation}\label{104-curv34}
t=\alpha H^{-1}
\end{equation}

which from $\alpha$ (\ref{98-curv28}) becomes

\begin{equation}\label{105-curv35}
t=\frac{-2n^2+3n-1}{n-2}H^{-1} ,
\end{equation}
so, the today value of the age of universe simply is $t=\alpha(n)
H_{0}^{-1}$

By a simple algorithm we can evaluate the age of universe taking
into account both the intervals and the $1\sigma$ range of
variability of the Hubble parameter, deduced from the supernovae
fit for each interval. We have considered as good predictions,
age estimates included between 10 and 18 Gyr. \\
First of all we have discarded the intervals of $n$ which give a
negative value of $t$. Obviously from Eq.(\ref{104-curv34}),
negative values for the age of the universe are obtained for
negative values of $\alpha$, so we have to exclude the ranges
$1/2<n<1$ and $n>2$ (Fig.5). The results are shown in Table 2.
\vspace{0.5cm}
\begin{table}
\begin{center}
 \begin{tabular}{|c|c|c|c|}
  \hline
  Range & $\Delta H(km\,sec^{-1}Mpc^{-1})$& $\Delta{n}$ & $t(n^{best})(Gyr)$ \\
  \hline
  $-100<n<1/2(1-\sqrt{3})$& $50-80$ & $-0.67\leq n<-0.37$ & $23.4$ \\ \hline
  $1/2(1-\sqrt{3})<n<1/2$ & $57-69$ & $-0.37<n \leq -0.07$ & $15.6$ \\ \hline
  $1<n<1/2(1+\sqrt{3})$ & $56-70$ & $1.28\leq n<1.36$ & $15.3$ \\ \hline
  $1/2(1+\sqrt{3})<n<2$ & $54-78$ & $1.37<n\leq 1.43$ & $24.6$ \\ \hline
\end{tabular}
\end{center}
\caption{\small The results of the age test. In the first column
is presented the tested range. Second column shows the $1\sigma$
$H_{0}$ range obtained by supernovae test, while in the third we
give the $n$ intervals, that is the values of $n$ which allow to
obtain ages of the universe comprised between 10 and 18 $Gyr$. In
the last column, the values of the age obtained for the best fit
value of each interval are reported.}
\end{table}
\vspace{1cm}
\\
\\
 As we have said above this test gives
interesting results.\\
The last check for our model is to verify if the range of $n$
significant for the data provides also an accelerated rate of
expansion. This test can be easily performed starting by relations
(\ref{nota}), (\ref{91-curv21}) and (\ref{98-curv28}). To have an
accelerated behaviour, the scale factor function
$a=a_{0}t^{\alpha}$ has to get values of $\alpha$ negative or
positive greater than one. We obtain that intervals $-0.67\leq n
\leq 0.37$ and $1.37\leq n \leq 1.43$ provide a negative
deceleration parameter with $\alpha>1$, that is cosmological
models expanding with an accelerated rate. Conversely the other
two intervals of Table 2 do not give interesting cosmological
dynamics in this sense ($q_{0}>0, 0<\alpha<1$).\\
Finally we are able to state that cosmological models based on a
relativistic lagrangian with a generic power of $R$ (different
from one) are physically coherent with observations.\\
This results indicate again that quintessence can be achieved
without scalar fields starting from effective fundamental
theories.

\section{\large\bf Third  hypothesis: Torsion quintessence}

\subsection{\normalsize\bf The model}

 Taking into account torsion  is a straightforward
generalization  to implement concepts as spin in General
Relativity \cite{hehl,trautman}. However, it was soon evident that
torsion, as considered e.g. in Einstein-Cartan-Sciama-Kibble
(ECKS) theory, does not seems to give relevant effects  in the
observed astrophysical structures. Nevertheless it was found that
for densities of the order of $10^{47} g/cm^3$ for electrons and
$10^{54} g/cm^3$ for protons and neutrons, torsion could give
observable consequences if all the spins of the particles are
aligned. These huge densities can be reached only in the early
universe so that cosmology is the only viable approach to test
torsion effects \cite{desabbata1}. However no relevant tests
confirming the presence of torsion have been found until now and
it is still an open debate if the space-time is Riemannian or not.
Considering the cosmological point of view and, in particular the
primordial phase transitions and inflation
\cite{kolb,peebles,desabbata2},  it seems very likely that, in
some regions of the early universe, the presence of local magnetic
fields could have aligned  the spins of particles. At very high
densities, this effect could influence the evolution of primordial
perturbations remaining as an imprint in today observed large
scale structures. In other words, a main goal could be to select
perturbation scales connected to the presence of torsion in early
epochs which give today-observable cosmological effects
\cite{cosimo}.

From another point of view, the presence of torsion could give
observable effects without taking into account clustered matter.
Here, we want to investigate if the quintessential scheme can be
achieved by taking into account  theories of gravity with torsion
\cite{captor}.

In the ECSK theory, the affine connection is non-symmetric in its
lower indices and the antisymmetric part

\beq\label{106-tor1} \Gamma^a_{[bc]}=S^{\phantom{bc}a}_{bc} \eeq

\noindent is called torsion. Usually, we can divide such an
antisymmetric part into three components:  one of them is
 irreducible,
while the other two can be set to zero \cite{goenner,classtor}.
This assumption yields the simplest theory containing torsion.

Furthermore, we can express by a 4-vector
\beq \label{107-tor2}
\sigma^a=\epsilon^{abcd}S_{bcd} \eeq

\noindent the totally antisymmetric part of torsion. If one
imposes to it the symmetries of a background which is homogeneous
and isotropic, it follows that, in comoving coordinates, only the
component $\sigma^0$ survives as a function depending only on
cosmic time (see \cite{goenner,classtor,tsamparlis}).

For a perfect fluid, the Einstein-Friedmann cosmological equations
in presence of torsion can be written as usual

\beq\label{108-tor3} \frac{\ddot{a}}{a}=-\frac{1}{6}(\tilde{\rho}+3\tilde{p})\,,
\eeq

\noindent and

 \beq\label{109-tor4} \left(\frac{\dot{a}}{a}\right)^{2}+\frac{k}{a^{2}}=
\frac{1}{3}\tilde{\rho}\,. \eeq

\noindent Energy density and pressure can be assumed in the forms
 \cite{goenner, classtor}

\beq \label{110-tor5} \tilde\rho=\rho +  f^2\,, \qquad \tilde p= p
-f^2\,, \eeq

where $f$ is a function related to $\sigma^0$, while $\rho$ and
$p$ are the usual quantities of General Relativity. This choice
can be pursued since
  we can define $S_{abc}=S_{[abc]}=f(t)$ where $f(t)$ is a
 generic function of time which we consider as the source of torsion.
 For a detailed discussion of this point see \cite{classtor}.

As usual, we can define a stress-energy tensor of the form \beq
\label{111-tor6}
T^{tot}_{ab}=(\tilde{p}+\tilde{\rho})u_{a}u_{b}-\tilde{p}g_{ab}\,,
\eeq which, by Eqs. (\ref{110-tor5}), can be splitted as
\beq\label{112-tor7}
 T^{tot}_{ab}=T^{M}_{ab}+T^{torsion}_{ab}\,.
\eeq Due to the contracted Bianchi identity, we have \beq
\label{113-tor8} T^{tot\,;b}_{ab}=0\;; \eeq from which, we can
assume that (cfr. \cite{minkowski}) \beq \label{114-tor9}
T^{M;b}_{ab}=0\,,\qquad T^{torsion\,;b}_{ab}=0\,. \eeq In the FRW
space-time, Eq.(\ref{113-tor8}) becomes
 \beq\label{115-tor10}
\dot{\tilde{\rho}}+3H(\tilde\rho + \tilde p)=0 \eeq which is \beq
\label{116-tor11} \dot{\rho} + 3H(\rho +  p) = -2f\dot{f}\,. \eeq
From Eqs.(\ref{114-tor9}), both sides of (\ref{116-tor11}) vanish
independently so that \beq\label{117-tor12}
f(t)=f_{0}=\mbox{constant.} \eeq

In other worlds, a torsion field gives rise to a constant energy
density, i.e. a ${\it torsion}-\Lambda$ term. Taking into account
standard matter which equation of state is defined into the above
Zeldovich range, we obtain

\beq \label{118-tor13}
 \tilde{\rho}=\rho_{0}\left[\frac{a_0}{a}
\right]^{3(\gamma+1)}+f_{0}^2\,. \eeq

Inserting this result into the cosmological equations, we get, in
any case,  a monotonic expansion being $f_{0}^2>0$, $\dot{a}^2>0$.
The condition to obtain the accelerated behaviour is \beq
\label{119-tor14} \rho+3p<2f_{0}^2\,, \eeq so that acceleration
depends on the torsion density.

In a dust-dominated universe, we have \beq \label{120-tor15}
\tilde{\rho}=\rho_{0}\left(\frac{a_{0}}{a}\right)^{3}+f_{0}^{2}\,,
\qquad \tilde{p}=-f_{0}^2\,,\eeq and then solving the cosmological
equations we obtain the general solution:

\beq \label{121-tor16}
a(t)=\left(\frac{a_0^3\rho_0}{2f_0}\right)^{1/3}
\left[\cosh(f_{0}t)-1\right]^{1/3} \,. \eeq

Obviously, if $f_{0}t\rightarrow 0$, we have $\displaystyle{\cosh
(f_{0}t)-1\simeq (f_{0}t)^2}$ and then $a\sim t^{2/3}$, as it has
to be.

This results tell us that a sort of dark energy (a
torsion-$\Lambda$ term) can be obtained without considering
additional scalar fields in the dynamics but only assuming that
the space-time is $U_4$ instead of $V_4$. A preminent role is
played by the type of torsion we are going to consider
\cite{classtor}. However, such a density should be comparable to
the observed limits of dark energy (\ie $\Omega_{\Lambda}\sim
0.65\div 0.7$) in order to give relevant effects. Another point is
that such a {\it torsion quintessence} should match the issues of
cosmic coincidence
 \cite{steinhardt} as scalar field quintessence. This point strictly depends
on  the type of torsion since torsion can be or not  related to
the spin density \cite{classtor}. In the first case, the spin of
baryonic and non-baryonic matter would rule the dark energy
(torsion) density.

\subsection{\normalsize\bf Matching Torsion quintessence with
observations}

\subsubsection{\normalsize\bf The Supernovae SNe Ia method}

As in previous cases we compare the theoretical estimates of
luminosity distance with the results obtained for type Ia
supernovae. Our analysis is based again on the sample data
\cite{wang} from SCP and HZT observations
\cite{perlmutter,riess}\\

The luminosity distance in the model we are considering is
completely equivalent to the one in a spatially flat universe with
a non-zero cosmological constant. Thus $d_L(z)$ is simply given
as\,:

\begin{equation}
d_L(z) = (1 + z) \int_{0}^{z}{dz' [ \Omega_M (1 + z')^3 +
\Omega_{tor} ]^{-1/2}} \ . \label{eq: dl}
\end{equation}
where $\Omega_ {tor} = 1 - \Omega_M$ plays the same role as the
usual $\Omega_{\Lambda}$.

Our model can be, so, fully characterized by two parameters\,: the
today Hubble constant $H_0$ and the matter density $\Omega_M$. We
find their best fit values minimizing the $\chi^2$ defined in this
case as\,:

\begin{equation}
\chi^2(H_0, \Omega_M) = \sum_{i}{\frac{[\mu_i^{theor}(z_i | H_0,
\Omega_M) - \mu_i^{obs}]^2} {\sigma_ {\mu_{0},i}^{2} +
\sigma_{mz,i}^{2}}} \label{eq: defchi}
\end{equation}
where the sum is over the data points \cite{wang}.\\
We remember that the $\sigma_{\mu_{0}}$ and $\sigma_{m_{z}}$ are
the errors related to the distance modulus evaluation, their
estimate is done as in the previous cases \cite{wang}.\\
The results of the fit are presented in Fig.6 where we show the
1,2 and 3\,$\sigma$ confidence regions in the $(\Omega_M, H_0)$
plane.

\begin{figure*}[ht]
\centering \resizebox{10cm}{!}{\includegraphics{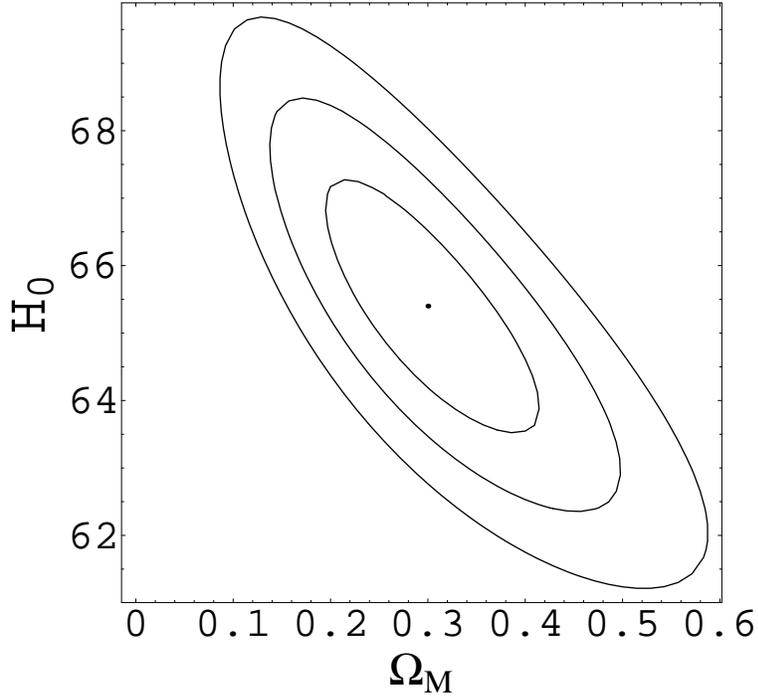}}
\hfill \caption{\small 1, 2 and 3$\sigma$ confidence regions in
the $(\Omega_M, H_0)$ plane. The central dot represents the best
fit values\,: $\Omega_M = 0.3, H_0 = 65.4 \ {\rm km \ s^{-1} \
Mpc^{-1}}$\,,}
\end{figure*}

The best fit values (with $1 \sigma$ error) turn out to be\,:

\begin{displaymath}
\Omega_M = 0.30 \pm 0.08 \ \ , \ \ H_0 = 65.4 \pm 1.2 \ km \
s^{-1} \ Mpc^{-1} \ .
\end{displaymath}
which allow to conclude that a torsion $\Lambda$-term could
explain observation very well. On the other hand, we can estimate
the torsion density contributions which result  to be\,:

\begin{displaymath}
f_0^2 = (5.6 \pm 0.7) {\times} 10^{-30} \ g \ cm^{-3} \,,
\end{displaymath}
which is a good value if compared to the cosmological critical
density. It is worthwhile to note that, in the case of SNeIa, the
error on $H_0$ does not take into account systematic uncertainties
due to possible calibration errors.

\subsubsection{\normalsize\bf The Sunyaev-Zeldovich/X-ray method}

Besides the above results,  we can discuss how the Hubble constant
$H_0$ and the torsion density parameter $\Omega_{\it tor}$ can be
constrained also by the angular diameter distance $D_A$ as
measured using the Sunyaev-Zeldovich effect (SZE) and the thermal
bremsstrahlung (X-ray brightness data) for galaxy clusters. \\
We will limit our analysis to the so called {\it thermal} or {\it
static} SZE, which is present in all the clusters, neglecting the
${\it kinematic}$ effect, which is present in those clusters with
a nonzero peculiar velocity with respect to the Hubble flow along
the line of sight. Typically the thermal SZE is an order of
magnitude larger than the kinematic one.
 The shift of temperature is:
\begin{equation}
\frac{\Delta T}{T_0} = y\left[ x \, \mbox{coth}\,
\left(\frac{x}{2} \right) -4 \right], \label{eq:sze5}
\end{equation}
where ${\displaystyle x=\frac{h \nu}{k_B T}}$ is a dimensionless
variable, $T$ is the radiation temperature, and $y$ is the so
called Compton parameter, defined as the optical depth $\tau =
\sigma_T \int n_e dl$ times the energy gain per scattering:
\begin{equation}\label{compt}
  y=\int  \frac{K_B T_e}{m_e c^2} n_e \sigma_T dl.
\end{equation}
In Eq.~(\ref{compt}), $T_e$ is the temperature of the electrons in
the intracluster gas, $m_e$ is the electron mass, $n_e$ is the
numerical density of electrons, and $\sigma_T$ is the  cross
section of Thompson electron scattering. We have used the
condition $T_e \gg T$ ($T_e$ is of the order $10^7\, K$ and $T$,
which is the CMBR temperature is $\simeq 2.7K$). Considering the
low frequency regime of the Rayleigh-Jeans approximation, we
obtain
\begin{equation}
 \frac{\Delta T_{RJ}}{T_0}\simeq -2y
 \label{eq:sze5bis}
\end{equation}
The next step  to quantify the SZE decrement is that we need to
specify the models for the intracluster electron density and
temperature distribution. The most commonly used model is the so
called isothermal $\beta$ model  \cite{cavaliere}. We have
\begin{eqnarray}
& & n_e (r) = n_e (r) = n_{e_0} \left( 1 + \left(
\frac{r}{r_e} \right)^2 \right)^{-\frac{3 \beta}{2}}\,, \\
& & T_e (r) = T_{e_0}~, \label{eq:sze6}
\end{eqnarray}
being $n_{e_0}$ and $T_{e_0}$, respectively the central electron
number density and temperature of the intracluster electron gas,
$r_e$ and $\beta$ are fitting parameters connected with the model
  \cite{sarazin}. For the effect of cluster modelling see \cite{jetzer}.
From Eq. (\ref{compt}) we have
\begin{equation}
\frac{\Delta T}{T_0} = -\left(\frac{2 K_B \sigma_T T_{e_0} \,
n_{e_0}}{m_e c^2}\right) \,\, \Sigma \,,\label{eq:sze7}
\end{equation}
being
\begin{equation}
\Sigma = \int^\infty_0 \left( 1 + \left( \frac{r}{r_c}
\right)^2\right)^{-\frac{3 \beta}{2}} dr\,. \label{eq:sze8}
\end{equation}
The integral in Eq.~(\ref{eq:sze8}) is overestimated  since
clusters have a finite radius. The effects of the finite extension
of the cluters are analyzed in~\cite{jetzer,cooray98b}.

A simple geometrical argument converts the integral in
Eq.~(\ref{eq:sze8}) in angular form, by introducing the angular
diameter distance, so that

\begin{equation}
\Sigma = \theta_c \left(1 + \left(
\frac{\theta}{\theta_2}\right)^2 \right)^{1/2 - 3 \beta/2}
\sqrt{\pi} \, \frac{\Gamma \left( \frac{3 \beta}{2} -
\frac{1}{2}\right)}{\Gamma \left( \frac{3 \beta}{2} \right)} \,
r_{DR}. \label{eq:sze9}
\end{equation}

In terms of the dimensionless angular diameter distances, $d_A$
(such that $D_A=\displaystyle\frac{c}{H_0} d_A$) we get
\begin{equation}
\frac{\Delta T (\theta)}{T_0} = -
\frac{2}{H_0}\left(\frac{\sigma_T K_B T_{ec}n_{e_0}}{m_e c}
\right)\sqrt{\pi}  \frac{\Gamma \left( \frac{3 \beta}{2}
\frac{1}{2} \right)}{\Gamma \left( \frac{3\beta}{2}\right)} \left(
1 - \left( \frac{\theta}{\theta_2}\right)^2 \right)^{\frac{1}{2}
(1 -3 \beta)} d_A, \label{eq:sze10}
\end{equation}
and, consequently, for the central temperature decrement, we get
\begin{equation}\label{eq:sze10bis}
\frac{\Delta T (\theta =0)}{T_0}=-
\frac{2}{H_0}\left(\frac{\sigma_T K_B T_{ec}n_{e_0}}{m_e c}\right)
\sqrt{\pi} \frac{\Gamma \left( \frac{3 \beta}{2} \frac{1}{2}
\right)}{\Gamma \left( \frac{3\beta}{2}\right)}\frac{c}{H_0} d_A.
\end{equation}
The factor $\displaystyle \frac{c}{H_0} d_A$  in
Eq.~(\ref{eq:sze10bis}) carrys the dependence on the thermal SZE
on both the cosmological models (through $H_0$ and the Dyer-Roeder
distance $d_A$) and the redshift (through $d_A$). From
Eq.~(\ref{eq:sze10bis}), we can note that the central electron
number density is proportional to the inverse of the angular
diameter distance, when calculated through SZE measurements. This
circumstance allows to determine the distance of cluster, and then
the Hubble constant, by the measurements of its thermal SZE and
its X-ray emission.

This possibility is based on the different power laws, according
to which  the decrement of the temperature in  SZE,
$\displaystyle\frac{\Delta T(\theta =0)}{T_0}$ , and X-ray
emissivity, $S_X$, scale with respect to the electron density. In
fact, as above pointed out, the electron density, when calculated
from SZE data, scales as $d^{-1}_{A}$~( $n^{SZE}_{e0}\propto
d^{-1}_A$), while the same one scales as
$d^{-2}_{A}$~($n^{X-ray}_{e0}\propto d^{-2}_A$) when calculated
from X-ray data. Actually, for the X-ray surface brightness,
$S_X$,  assuming for the temperature distribution of $T_e=T_{e0}$,
we get the following formula:
\begin{equation}\label{sx}
 S_X=\frac{\epsilon_X}{4\pi}{n_{e0}^2}\frac{1}{ (1+z)^3} \theta_c \frac{c}{H_0} d_A I_{SX},
\end{equation}
being \[I_{Sx}=\int^{\infty}_0 \left(\frac{n_e}{n_{e0}}\right)^2
dl\,,\] the X-ray structure integral, and $\epsilon_X$ the
spectral emissivity of the gas (which, for $T_e\geq
3{\times}10^{7}$, can be approximated by a typical value:
$\epsilon_X = \epsilon \sqrt{T_e}$, , with $\epsilon \simeq 3.0
{\times} 10^{-27} n_p^2$ erg $cm^{-3}$ $s^{-1}$
$K^{-1}$~\cite{sarazin}) . The angular diameter distance can be
deduced by eliminating the electron density from
Eqs.~(\ref{eq:sze10bis}) and (\ref{sx}), yielding:

\begin{equation}
\frac{y^2}{S_X}= \frac{4 \pi (1+z)^3}{\epsilon} {\times}
 \left(\displaystyle\frac{k_B \sigma_{T}}{m_e c^2}\right)^2 {T_{e0}}^{3/2}
 \theta_c \frac{c}{H_0} d_A {\times}
\frac{\left[B(\frac{3}{2}\beta-\frac{1}{2},
\frac{1}{2})\right]^2}{B(3\beta-\frac{1}{2}, \frac{1}{2})}\,\,,
\label{eq:sze11}
\end{equation}
where $B(a,b)=\displaystyle\frac{\Gamma(a)\Gamma(b)}{\Gamma(a+b)}$
is the Beta function.

It turns out that
\begin{equation}\label{eq:sze11bis}
D_A=\frac{c}{H_0}d_A\propto \frac{(\Delta T_0)^2}{S_{X0}
T^2_{e0}}\frac{1}{\theta_c},
\end{equation}
where all these quantities are evaluated along the line of sight
towards the cluster center (subscript 0), and $\theta_c$ is
referred to a characteristic cluster scale along the line of
sight. It is evident that the specific meaning of this scale
depends on the density model adopted for the cluster. In our
calculations we are using the so called $\beta$ model.

Eqs.~(\ref{eq:sze11}) allows  to  compute the Hubble constant
$H_0$, once the redshift of cluster is known and the other
cosmological parameters are, in same way, constrained. Since the
dimensionless Dyer-Roeder distance, $d_A$,  depends on
$\Omega_{{\it tor}}$, $\Omega_{M}$, comparing the  estimated
values with the theoretical formulas for $D_A$, it is possible to
obtain information about $\Omega_M$ $\Omega_{{\it tor}}$, and
$H_0$. Recently, distances of 18 clusters with redshift ranging
from $z\sim 0.14$ to $z\sim 0.78$ have been determined from a
likelihood joint analysis of SZE and X-ray observations (see
\cite{reese} and reference therein). Modeling the intracluster gas
as a spherical isothermal $\beta $-model allows to obtain
constraints on the Hubble constant $H_0$ in a standard
$\Lambda$-FRW model. We perform a similar analysis using angular
diameter distances measurements for a sample of 44 clusters,
constituted by the 18 above quoted  clusters and other 24 already
know data (see \cite{birk}).

 As indicated in
\cite{birk, reese},  the errors $\sigma$ are  only of statistical
nature. Taking into account our  model with torsion
(\ref{121-tor16}), the theoretical expression for the angular
diameter distances $D_A$ is
\begin{equation}\label{eq:szetors1}
  D_A(z)=\frac{1}{(1+z)^2}d_L(z) = \frac{1}{1+z} \int_{0}^{z}{dz' [ \Omega_M (1 +
z')^3 + \Omega_{tor} ]^{-1/2}}\,.
\end{equation}
We find the best fit values for $ \Omega_{tor}$ and $H_0$,
minimizing the reduced  $\chi^2$:
\begin{equation}
\chi^2(H_0, \Omega_M) = \sum_{i}{\frac{[{D_A}^{theor}(z_i | H_0,
\Omega_M) - {D_A}_i^{obs}]^2}{\sigma_ {D_A}^{2}}}.
\label{eq:eq:szetors2}
\end{equation}

The best fit values (at $1 \sigma$) turn out to be\,:
\begin{displaymath}
\Omega_M = 0.30 \pm 0.3 \ \ , \ \ H_0 = 68\pm 6 \ km \ s^{-1} \
Mpc^{-1} \ .
\end{displaymath}

in good agreement with the above fit derived from SNe Ia data.\\
\\

In conclusion, we have shown that the net effect of torsion is the
introduction of an extra-term into fluid matter density and
pressure which is capable of giving rise to an accelerated
behaviour of cosmic fluid. Being such a term a constant, we can
consider it a sort of torsion $\Lambda$-term. If the standard
fluid matter is dust, we can exactly solve dynamics which is in
agreement with the usual Friedmann model (to be precise
Einstein-de Sitter) as soon as torsion contribution approaches to
zero.

The next step has been to compare the result with observations in
order to see if such a torsion cosmology gives rise to a coherent
picture. We have used SNe Ia data, Sunyaev-Zeldovich effect and
X-ray emission from galaxy clusters. Using our model, we are
capable to reproduce the best fit values of $H_0$ and $\Omega_M$
which gives a cosmological model dominated by a cosmological
$\Lambda$-term. In other words, it seems that introducing torsion
(and then spins) in dynamics allows to explain in a {\it natural}
way the presence of cosmological constant or  a generic form of
dark energy without the introduction of exotic scalar fields.
Besides, observations allows to estimate torsion density which can
be comparable to other forms of matter energy $(\sim 5.5{\times}
10^{-30}\,g\,cm^{-3})$.

However, we have to say that we used only a particular form of
torsion and the argument can be more general if extended to all
the forms of torsion \cite{classtor}. Furthermore, being in our
case the torsion contribution a constant density, it is not
possible to solve {\it coincidence} and {\it fine tuning}
problems. To address these issues we need a form of torsion
evolving with time.

\section{\normalsize\bf Discussion, conclusions and perspectives}

In this review paper, we have discussed, from theoretical and
observational points of view, some approaches by which it is
possible to obtain the observed accelerated behaviour of cosmic
fluid recently reported by several authors.\\
Our point of view is, in some sense, conservative since we want to
investigate if and when such a behaviour can be achieved either
using fundamental effective theories of gravity or by correcting
standard perfect fluid cosmology as in the case of Van der
Waals.\\
The underlining philosophy is related to the fact that
quintessence is by now become a sort of paradigm (like inflation!)
so that every kind of scalar field (and potential) could be safely
introduced into the game also without any fundamental physics
motivation. Such a way of thinking is disturbing since we cannot
claim to solve a new problem by introducing new fundamental
ingredients.\\
With this premise in mind, we have analyzed three schemes
essentially based on "already known" aspects of fundamental physics.\\
The first is just a slight correction: what happens if, instead of
perfect fluids we adopt Van der Waals fluids as source in
Einstein-Friedmann cosmological equations?\\
This means to take into account a more realistic description of
the matter content of the universe, capable, on the other hand, of
implementing dynamics of cosmic flow where phase transitions are
enclosed. Van der Waals fluid equation can be read as a second
order correction (in terms of matter-energy density) of perfect
fluid equation.\\
It is interesting to see that for wide ranges of parameters
accelerated behaviours of scale factor can be achieved. Besides,
we get interesting matches with observational data, in particular
the Hubble-SneIa diagram and the age of the universe.\\
This is not a full self-consistent model but it is a useful
indication of the fact that observations can be explained using
more realistic cosmological fluid sources.\\
The curvature quintessence approach is completely different. Our
starting point is the fact that effective gravitational
lagrangians turn out to be extremely useful to study the quantum
behaviour of gravity. Such a formalism is based on the
introduction of higher-order corrections of curvature invariants
in the Einstein-Hilbert gravitational action in order to get
renormalization, at least, at a finite number of loops.\\
This scheme has given rise to several interesting inflationary
models (e.g. Starobinsky one), in which shortcomings of standard
cosmological model were cured in the framework of such geometrical
corrections. We asked for a similar approach at today very low
energies. In other words, if remnants of primordial quantum
gravity effects have survived until now, they could be responsible
of the observed unclustered dark energy and explain the
accelerated behaviour.  We worked out some fourth-order gravity
cosmological models and found that also in this case
quintessential issues can be matched. The comparison with
observation give suitable values of $H_{0}$ and of the age of
universe.\\
The last scheme investigated is a cosmological model with torsion.
Introducing spin into dynamic as further source of gravitational
field equation is the main goal of theories with torsion. We
wonder whether some kind of torsion field, among the several ones
existing, can work in order to give rise to accelerated behaviours
of cosmic fluid. Our results confirms such hypothesis and match
extremely well with observational data as discussed above.\\
In conclusion we have proposed alternative but physically founded
approaches to quintessence in order to see if such a paradigm can
be recovered in the framework of fundamental physics.\\
Obviously the presented models have to be improved in order to
match all the quintessential issues as the coincidence problem,
the low value of (dark energy) cosmological constant and the
compatibility with large scale structure.

\section*{\normalsize\bf Acknowledgement}
The authors are grateful to their friend V.F. Cardone, for the
useful discussions and comments on the topics.

\end{document}